\documentclass[pdftex,twocolumn,epjc3]{svjour3}
\RequirePackage[T1]{fontenc}
\smartqed  %
\RequirePackage{graphicx}
\RequirePackage{mathptmx}      %
\RequirePackage{flushend}
\RequirePackage[numbers,sort&compress]{natbib}
\RequirePackage[colorlinks,citecolor=blue,urlcolor=blue,linkcolor=blue]{hyperref}
\RequirePackage{amsmath} %
\RequirePackage{amssymb} %
\RequirePackage{color}
\RequirePackage{comment}
\RequirePackage{multirow}
\RequirePackage{rotating} %
\RequirePackage[english]{babel}
\RequirePackage{stfloats}
\RequirePackage{bm}
\journalname{Eur. Phys. J. C}

\begin{document}

\title{Information and treatment of unknown correlations\\
in the combination of measurements using the BLUE method}
\author{Andrea Valassi \thanksref{e1,addr1}
\and Roberto Chierici\thanksref{e2,addr2}}
\thankstext{e1}{e-mail: andrea.valassi@cern.ch}
\thankstext{e2}{e-mail: roberto.chierici@cern.ch}
\institute{CERN, Information Technology Department, CH-1211 Geneva 23, Switzerland\label{addr1}
\and CNRS, Institut de Physique Nucl\'eaire de Lyon, F-69622 Villeurbanne Cedex, France\label{addr2}}
\date{}
\maketitle

\begin{abstract}
We discuss the effect of large positive correlations in the combinations 
of several measurements of a single physical quantity 
using the Best Linear Unbiased Estimate (BLUE) method.
We suggest a new approach for comparing the relative 
weights of the different measurements in their contributions
to the combined knowledge about the unknown parameter,
using the well-established concept of Fisher information.
We argue, in particular, that one contribution to information comes  %
from the collective interplay of the measurements through their correlations
and that this contribution cannot be attributed to any of the 
individual measurements alone.
We show that negative coefficients in the BLUE weighted average
invariably indicate the presence of a regime of high correlations,
where the effect of further increasing some of these correlations
is that of reducing the error on the combined estimate.
In these regimes, we stress that 
assuming fully correlated systematic uncertainties
is not a truly conservative choice,
and that the correlations provided as input
to BLUE combinations need to be assessed with extreme care instead.
In situations where the precise evaluation of these correlations is impractical,
or even impossible, we provide tools to help experimental physicists 
perform more conservative combinations.
\keywords{correlated measurements \and estimation theory \and Fisher information \and best linear unbiased estimate}
\PACS{02.50.Sk %
\and 29.85.-b %
\and 14.65.Ha} %
\end{abstract}

\section{Introduction}
\label{sec:introduction}

Our knowledge about some of the most fundamental parameters of 
physics is derived from a vast number of measurements
produced by different experiments
using several complementary techniques.
Many statistical methods are routinely used~\cite{bib:pdg} 
to combine the available data
and extract the most appropriate estimates 
of the values and uncertainties for these parameters, properly taking into account 
all correlations between the measurements.
One the most popular methods for performing these combinations 
is the Best Linear Unbiased Estimate (BLUE) technique, 
an approach first introduced in the 1930's~\cite{bib:aitken} 
and whose reformulation in the context 
of high-energy physics~\cite{bib:lyons,bib:valassi} %
has been routinely used for the combination 
of the precision measurements performed by experiments 
at the LEP~\cite{bib:lep_ewwg}, Tevatron~\cite{bib:tev_top}
and LHC~\cite{bib:lhc_top} colliders, 
as well as in other domains.

\begin{sloppypar}
To quantify the ``relative importance'' of each measurement in its contribution 
to the combined knowledge about the measured physical quantity,
its coefficient in the BLUE weighted average is traditionally used.
In many examples in the literature where the BLUE technique has been used,
the combinations are dominated by systematic uncertainties,
often assumed as fully correlated among different measurements.
This often leads to situations where one or more measurements 
contribute with a negative BLUE coefficient,
pushing experimentalists to redefine the ``relative importance'' 
of a measurement as the absolute value of its BLUE coefficient, normalised to the sum of 
the absolute values of all coefficients~\cite{bib:tev_top, bib:lhc_top}.
In our opinion, this approach is \mbox{incorrect}.
\end{sloppypar}

In this paper, we propose a different approach for comparing the relative contributions 
of the measurements to the combined knowledge about the unknown parameter, 
using the well-established concept of Fisher information~\cite{bib:fisher}.
We also show that negative coefficients in the BLUE weighted average
invariably indicate the presence of very high correlations,
whose marginal effect is that of reducing the error %
on the combined estimate, rather than increasing it. 
In these regimes, we stress that taking systematic uncertainties
to be fully (i.e. 100\%) correlated is not a conservative assumption,
and we therefore argue that the correlations provided as inputs 
to BLUE combinations need to be assessed with extreme care.
In those situations where their precise evaluation is impossible, 
we offer a few guidelines and tools 
for critically re-evaluating these correlations,
in order to help experimental physicists perform more ``conservative'' combinations.
In our discussion, we will generally limit ourselves 
to BLUE combinations of a single measured parameter 
and where the correlations
used as inputs to the combination are positive.
Many of the concepts and tools we present
could be applied also to the more general cases
of BLUE combinations of several measured parameters,
and/or involving also negative correlations between measurements,
but this discussion is beyond the scope of this paper.

The outline of this article is the following.
In Sec.~\ref{sec:ri} we review the definition
of ``relative importance'' of a measurement in a BLUE combination
as presented by some papers in the literature
and we present our objections to it
by using a simple numerical example.
We then present our alternative definitions of %
information weights in Sec.~\ref{sec:information},
after a brief recall of the definition of Fisher information 
and of its relevant features.
By studying marginal information and information derivatives,
in Sec.~\ref{sec:highcorr}
we show that negative BLUE coefficients 
in the combination of several measurements
of one parameter are always a sign %
of a ``high-correlation'' regime, 
thus generalising the results presented
for two measurements
by the authors of Ref.~\cite{bib:lyons}.
In Sec.~\ref{sec:conservativeness} 
we go on to discuss practical guidelines and tools,
illustrated by numerical examples,
to identify correlations that may have been overestimated
and to review them in a more ``conservative'' way.
In Sec.~\ref{sec:conclusions}
we summarize our discussion
and present some concluding remarks.

\section{``Relative importance'' and negative BLUE coefficients}
\label{sec:ri}

\newcommand{\transpose}[1]{\tilde{#1}}
\newcommand{\yexp}{y} %
\newcommand{\ytru}{Y}
\newcommand{\yhat}{\hat{Y}}
\newcommand{\myvec}[1]{\mathbf{#1}}
\newcommand{\yexpvec}{\myvec{\yexp}}
\newcommand{\Uvec}{\myvec{U}}
\newcommand{\lvec}{\bm{\lambda}}
\newcommand{\lvect}{\bm{\transpose{\lambda}}}
\newcommand{\ofi}{_i}
\newcommand{\ofj}{_j}
\newcommand{\ofij}{_{ij}}
\newcommand{\ofji}{_{ji}}
\newcommand{\ofii}{_{ii}}
\newcommand{\ofjj}{_{jj}}
\newcommand{\ofik}{_{ik}}
\newcommand{\ofki}{_{ki}}
\newcommand{\ofjk}{_{jk}}
\newcommand{\ofkj}{_{kj}}
\newcommand{\nexp}{n}
\newcommand{\Mcov}{\mathcal{M}}
\newcommand{\Minv}{\mathcal{M}^{-1}}
\newcommand{\mymat}[1]{#1}
\newcommand{\McovMat}{\mymat{\mathcal{M}}}
\newcommand{\MinvMat}{\mymat{\mathcal{M}}^{-1}}
\newcommand{\syhat}{\sigma_{\raisebox{1pt}{\tiny$\hat{Y}$}}}
\newcommand{\ofA}{_{\raisebox{-1pt}{\scriptsize A}}}
\newcommand{\ofB}{_{\raisebox{-1pt}{\scriptsize B}}}
\newcommand{\ofBo}{_{\raisebox{-1pt}{\scriptsize B1}}}
\newcommand{\ofBoo}{_{\raisebox{-1pt}{\scriptsize B11}}}
\newcommand{\ofBot}{_{\raisebox{-1pt}{\scriptsize B12}}}
\newcommand{\ofBt}{_{\raisebox{-1pt}{\scriptsize B2}}}
\newcommand{\sA}{\sigma\ofA}
\newcommand{\sB}{\sigma\ofB}
\newcommand{\sBo}{\sigma\ofBo}
\newcommand{\sBt}{\sigma\ofBt}
\newcommand{\si}{\sigma_i}
\newcommand{\lA}{\lambda\ofA}
\newcommand{\lB}{\lambda\ofB}
\newcommand{\lBo}{\lambda\ofBo}
\newcommand{\lBt}{\lambda\ofBt}
\newcommand{\yA}{\yexp\ofA}
\newcommand{\yB}{\yexp\ofB}
\newcommand{\yBo}{\yexp\ofBo}
\newcommand{\yBoo}{\yexp\ofBoo}
\newcommand{\yBot}{\yexp\ofBot}
\newcommand{\yBt}{\yexp\ofBt}
\newcommand{\rBoBt}{\rho}
\newcommand{\sAS}{\sA^2}
\newcommand{\sBS}{\sB^2}
\newcommand{\myh}{}

\begin{sloppypar}
In the BLUE technique, the best linear unbiased estimate of each unknown parameter 
is built as a weighed average of all available measurements.
The coefficients multiplying the measurements in each linear combination
are determined as those that minimize its variance,
under the constraint of a normalisation condition which ensures that this represents 
an unbiased estimate of the corresponding parameter.
As discussed extensively in Refs.~\cite{bib:lyons,bib:valassi,bib:bos},
this technique is equivalent to minimizing the weighted sum of squared distances 
of the measurements from the combined estimates,
using as weighting matrix the input covariance matrix of the measurements,
which is assumed to be known a priori. 
\end{sloppypar}

\newcommand{\Exp}{{\mathbb E}}
In the case of $\nexp$ measurements $\yexp\ofi$ 
of a single parameter whose true value is $\ytru$, in particular,
the best linear unbiased estimate $\yhat$ can be determined as follows.
First, the BLUE should be a linear combination
$\yhat = \sum_{i=1}^{\nexp} \, \lambda\ofi \, \yexp\ofi $
of the available measurements.
Second, the BLUE should be an unbiased estimator,
i.e. its expectation value $\Exp[\yhat]$
should be equal to the true value $\ytru$ of the unknown parameter.
Assuming that each measurement
is also an unbiased estimator,
i.e. that its outcomes are distributed as random variables
with expectation values $\Exp[\yexp\ofi]=\ytru$,
this is equivalent to requiring a normalisation condition
$\sum_{i=1}^{\nexp} \lambda\ofi = 1$
for the coefficients $\lambda\ofi$ in the linear combination.
Third, the BLUE should be the best of
such unbiased linear combinations,
i.e. that for which the combined variance
$\syhat^2 = \sum_{i=1}^{\nexp} \sum_{j=1}^{\nexp} 
\lambda\ofi \, \Mcov\ofij \, \lambda\ofj$,
where $\McovMat$ is the covariance matrix of the measurements,
is minimized.
It is then easy to show~\cite{bib:lyons} 
that $\yhat$ is the best linear unbiased estimate
if the coefficients $\lambda\ofi$ are equal to
\begin{equation}
  \lambda\ofi = 
  \frac{\raisebox{2pt}{$\bigl(\MinvMat\Uvec\bigr)\ofi$}}
  {\raisebox{-2pt}{$\transpose{\Uvec}\MinvMat\Uvec$}}
  \myh,
  \label{eq:coeff1}
\end{equation}
where $\Uvec$ is a vector whose elements are all equal to 1.

While the normalisation condition 
ensures that the coefficients $\lambda\ofi$ sum up to 1,
one peculiar and somewhat counter-intuitive feature of this method is that
some of these individual coefficients may be negative.
Negative coefficients in the BLUE weighted averages 
apparently still pose a problem of interpretation sometimes,
especially if these coefficients are used to compare
the contributions of the different measurements
to the combined knowledge about the measured observable.
For instance,
the ``relative importance'' of each measurement 
in the combination of ATLAS and CMS results 
on the top quark mass~\cite{bib:lhc_top}
was defined as the absolute value of its coefficient in the BLUE weighted average, 
divided by the sum of the absolute values of the coefficients for all input measurements,
\newcommand{\ri}{\mathrm{RI}}
\begin{equation}
\ri\ofi = \frac{|\lambda\ofi|}{\sum_{j=1}^{\nexp} |\lambda\ofj|} 
\myh.
\label{eq:ri}
\end{equation}
The same procedure had already been used to visualize
the ``weight that each measurement carries in the combination'' 
of CDF and D0 results on the top quark mass~\cite{bib:tev_top}.
In both cases, 
the relative importances of the $\nexp$ measurements sum up to 1
by definition, $\sum_{i=1}^\nexp\ri\ofi = 1$.

In our opinion, this procedure
is an artefact that is conceptually wrong and suffers from two important limitations: 
first, it is not internally self-consistent and
may easily lead to numerical conclusions which go against common sense;
second, it does not help to understand in which way the results with negative coefficients 
contribute to reducing the uncertainties on the combined estimates.
We will use a simple example to illustrate the first objection.
Consider the combination of two uncorrelated measurements
$\yA=103.00\pm3.87$ and $\yB=98.00\pm3.16$
of an observable $\ytru$ in the appropriate units.
The covariance matrix is then
\begin{equation}
\renewcommand{\arraystretch}{1.2}
\left(\begin{array}{c|c} \sA^2 & 0\\ \hline 0 & \sB^2 \\ \end{array} \right)
=
\left(\begin{array}{c|c} 15.00 & 0\\ \hline 0 & 10.00 \\ \end{array} \right)
\myh
\renewcommand{\arraystretch}{1}
\end{equation}
and the BLUE for their combination is $\yhat=\lA\yA+\lB\yB=100.00\pm2.45$,
where the coefficients of these two uncorrelated measurements 
in the BLUE weighted average, $\lA\!=\!0.4$ and $\lB\!=\!0.6$,
are proportional to the inverses of the variances $\sA^2$ and $\sB^2$
as expected from simple error propagation.
It is rather intuitive in this case 
to claim that the relative contributions
to the knowledge about $\ytru$
contributed by the two independent measurements A and B 
can be quantified by their BLUE coefficients, 40\% for A and 60\% for B.
As $\lA$ and $\lB$ are both positive, these are also 
the ``relative importances'' of A and B according to~Eq.~\ref{eq:ri}.

Imagine now that $\yB$ is not the result of a direct measurement,
but is itself the result of the combination of two
measurements $\yBo=99.00\pm4.00$ and $\yBt=101.00\pm8.00$,
where a high positive correlation $\rBoBt=0.875$ between them
leads to negative BLUE coefficients in their weighted average
$\yB=1.5\!\times\!\yBo-0.5\!\times\!\yBt=98.00\pm3.16$.
Instead of combining first $\yBo$ and $\yBt$ and then adding $\yA$,
one could also combine $\yA$, $\yBo$ and $\yBt$ directly 
using the full covariance matrix
\begin{equation}
\renewcommand{\arraystretch}{1.2}
\left(\begin{array}{c|cc}
 \sA^2 & 0 & 0 \\ \hline 0 & \sBo^2 & \rBoBt\sBo\sBt \\ 0 & \rBoBt\sBo\sBt & \sBt^2 \\ 
\end{array}\right)
=
\left(\begin{array}{c|cc}
 15.00 & 0 & 0 \\ \hline 0 & 16.00 & 28.00 \\ 0 & 28.00 & 64.00 \\ 
\end{array}\right)
\myh.
\renewcommand{\arraystretch}{1}
\end{equation}
This yields $\yhat=\lA\yA+\lBo\yBo+\lBt\yBt=100.00\pm2.45$,
where the BLUE coefficients in this overall weighted average
are given by $\lA\!=\!0.4$, $\lBo\!=\!0.9$ and $\lBt\!=\!-0.3$.

As expected, the final numerical result for $\yhat$ is of course the same 
whether it is obtained from the combination of $\yA$ and $\yB$
or from the combination of $\yA$, $\yBo$ and $\yBt$.
It is also not surprising that the BLUE coefficient 
$\lA\!=\!0.4$ for $\yA$ is the same in both combinations,
as this is an independent measurement 
that is not correlated to either $\yBo$ or $\yBt$
(the sum of whose BLUE coefficients, $\lBo+\lBt=0.9\!-\!0.3\!=\!0.6$,
of course, equals the BLUE coefficient $\lB$ of $\yB$).
What is rather surprising, however,
is that the ``relative importance'' of $\yA$
computed using normalised absolute values of the BLUE coefficients
is very different in the two cases:
\begin{equation}
\renewcommand{\arraystretch}{1.2}
\left\{
\begin{array}{lcccc}
\ri_A \mathrm{(combining\ A,\ B)} & = & \frac{0.4}{0.4+(0.9-0.3)} & = & 40.0\% \myh, \\
\ri_A \mathrm{(combining\ A,\ B1,\ B2)} & = & \frac{0.4}{0.4+0.9+0.3} & = & 25.0\% \myh. \\
\end{array}
\right.
\renewcommand{\arraystretch}{1}
\end{equation}
In our opinion, this is an internal inconsistency of Eq.~\ref{eq:ri},
as common sense suggests that the relative contribution of $\yA$ 
to the knowledge about $\ytru$ is the same in both combinations.
In particular, we consider that the contribution of $\yA$ is indeed 40\%,
and that this is underestimated as 25\% in the second combination
because the relative contributions of $\yBo$ and $\yBt$
in the presence of negative BLUE coefficients 
are not being properly assessed and are overall overestimated.

More generally, the problem with defining
the ``relative importances'' of measurements according to~Eq.~\ref{eq:ri} is
that the coefficient with which a measurement enters 
the linear combination of all measurements in the BLUE,
i.e. its ``weight'' in the BLUE weighted average,
is being confused with the impact or ``weight''
of its relative contribution to the knowledge
about the measured observable.
In the following we will therefore clearly distinguish 
between these two categories of ``weights'':
we will sometimes refer to the BLUE coefficient $\lambda\ofi$
of a measurement as its ``central value weight''~(CVW),
while we will use the term ``information weight''~(IW)
to refer to, using the same words as in Refs.~\cite{bib:tev_top,bib:lhc_top},
its ``relative importance'' or the ``weight it carries in the combination''.
We will propose and discuss our definitions
of intrinsic and marginal information weights in the next section,
using the well-established concept of Fisher information.

\newcommand{\ofa}{_\alpha}
\newcommand{\ofb}{_\beta}
\newcommand{\ofab}{_{\alpha\beta}}
\newcommand{\ofaa}{_{\alpha\alpha}}
\newcommand{\ntru}{N}
\newcommand{\Ucal}{\mathcal{U}}
\newcommand{\Ucalt}{\transpose{\mathcal{U}}}
\newcommand{\UcalMat}{\mymat{\mathcal{U}}}
\newcommand{\UcaltMat}{\transpose{\mymat{\mathcal{U}}}}
\newcommand{\score}{s}
\newcommand{\xtru}{X} %
\newcommand{\xhat}{\hat{x}} %
\newcommand{\xtruvec}{\myvec{\xtru}}
\newcommand{\xhatvec}{\myvec{\xhat}}
\newcommand{\ofxtru}{\raisebox{-1pt}{$^{\scriptsize(\xtruvec)}$}}
\newcommand{\If}{\mathcal{I}}
\newcommand{\Ifx}{\mathcal{I}\ofxtru}
\newcommand{\IfMat}{\mymat{\mathcal{I}}}
\newcommand{\IfxMat}{\mymat{\mathcal{I}}\ofxtru}
\newcommand{\pdf}{p}
\newcommand{\sAst}{\sigma_{A_{stat}}}
\newcommand{\sAsy}{\sigma_{A_{sys}}}
\newcommand{\sBst}{\sigma_{B_{stat}}}
\newcommand{\sBsy}{\sigma_{B_{sys}}}
\newcommand{\sAstS}{\sAst^2}
\newcommand{\sAsyS}{\sAsy^2}
\newcommand{\sBstS}{\sBst^2}
\newcommand{\sBsyS}{\sBsy^2}
\newcommand{\rAB}{\rho_{AB}}
\newcommand{\rABsy}{\rho_{AB_{sys}}}
\newcommand{\rAC}{\rho_{AC}}
\newcommand{\rBC}{\rho_{BC}}
\newcommand{\rij}{\rho_{ij}}
\newcommand{\cov}{\mathrm{cov}}
\newcommand{\var}{\mathrm{var}}
\newcommand{\pb}{\mathrm{pb}}
\newcommand{\pbi}{\mathrm{pb}^{-1}}
\newcommand{\sC}{\sigma_{C}}
\newcommand{\sCcor}{\sigma_{C,\mathrm{cor}}}
\newcommand{\sDcor}{\sigma_{D,\mathrm{cor}}}
\newcommand{\nerr}{S}
\newcommand{\ierr}{s}
\newcommand{\ofierr}{^{\raisebox{0pt}{{\fontsize{5}{6}\selectfont$[$}\scriptsize$\ierr${\fontsize{5}{6}\selectfont$]$}}}}
\newcommand{\jerr}{s'}
\newcommand{\ofytru}{^{(\ytru)}}
\newcommand{\Info}{I}
\newcommand{\Ify}{\Info\ofytru}
\newcommand{\iiw}{\mathrm{IIW}}
\newcommand{\iiwcorr}{\iiw_{\mathrm{corr}}}

\section{Fisher information and ``information weights''}
\label{sec:information}

In this section, we present our definitions 
of intrinsic and marginal {information weights},
after briefly recalling the definition of Fisher information 
and summarizing its main relevant features.
A more general discussion of Fisher information
and its role in parameter estimation in experimental science 
is well beyond the scope of this paper and can be found 
in many textbooks on statistics
such as the two excellent reviews in Refs.~\cite{bib:bos,bib:james},
which will largely be the basis of the overview presented in this section.

\subsection{Definition of Fisher information}

Consider $\nexp$ experimental measurements
$\yexp\ofi\!=\!\{\yexp_1,\ldots\yexp_\nexp\}$
that we want to use to infer the true values 
$\xtru\ofa\!=\!\{\xtru_1,\ldots\xtru_\ntru\}$ 
of $\ntru$ unknown parameters, with $\nexp>\ntru$
(though each of the $\yexp\ofi$ need not necessarily be 
a direct measurement of one of the parameters $\xtru\ofa$).
We will use the symbols $\yexpvec$ and $\xtruvec$ to indicate the vectors 
of all $\yexp\ofi$ and of all $\xtru\ofa$, respectively.
The measurements $\yexpvec$ are random variables distributed 
according to a 
probability density function $\pdf(\yexpvec;\xtruvec)$ 
that is defined under the normalisation condition
$\int\pdf(\yexpvec;\xtruvec) d\yexp_1 \ldots d\yexp_\nexp = 1$.
The sensitivity of the measurements to the unknown parameters
can be represented by the Fisher ``score vector''
$\score\ofxtru\ofa = \partial\log\pdf(\yexpvec;\xtruvec)/\partial\xtru\ofa$,
which is itself a random variable, 
defined in the $\nexp$-dimensional space of the measurements
and whose value in general also depends on the parameters $\xtruvec$.
Under certain regularity conditions
(in summary, 
the ranges of values of $\yexpvec$ must be independent of $\xtruvec$,
and $\pdf(\yexpvec; \xtruvec)$ must be regular enough to allow
$\partial^2/\partial\xtru\ofa\xtru\ofb$ and $\int d\yexpvec$ to commute),
it can be shown~\cite{bib:bos,bib:james} that
the expectation value of the Fisher score is the null vector,
$\int\,\score\ofxtru\ofa \, \pdf(\yexpvec; \xtruvec) \, d\yexp_1 \ldots d\yexp_\nexp = 0$. %
The Fisher information matrix,
which in the following we will generally refer to simply as ``information'',
is defined as the covariance of the score vector:
as the expectation value of the score is null, %
this can simply be written as
\begin{multline}
\Ifx\ofab =
\Exp\bigl[\frac{\partial\log\pdf(\yexpvec;\xtruvec)}{\partial\xtru\ofa}
\,\frac{\partial\log\pdf(\yexpvec;\xtruvec)}{\partial\xtru\ofb}\bigr]\\
=
\int\,\frac{\partial\log\pdf(\yexpvec;\xtruvec)}{\partial\xtru\ofa}
\,\,\frac{\partial\log\pdf(\yexpvec;\xtruvec)}{\partial\xtru\ofb}
\,\,\pdf(\yexpvec;\xtruvec) \, d\yexp_1 \ldots d\yexp_\nexp %
\myh.
\label{eq:infodef}
\end{multline}
Information is thus defined as 
the result of an integral over $d\yexp_1 \ldots d\yexp_\nexp$ and 
does not depend on the specific numerical outcomes
of the measurements $\yexp\ofi$, although in general it is a function 
of the parameters $\xtru\ofa$ instead. In other words, 
information is a property of the measurement process,
and more particularly of the errors on the measurements 
and of the correlations between them, 
rather than of the specific measured central values $\yexp\ofi$.

As pointed out in Ref.~\cite{bib:james}, 
Fisher information is a valuable tool for assessing quantitatively 
the contribution of an individual measurement to our knowledge
about an unknown parameter inferred from it, 
because it possesses three remarkable properties.

First, information increases with the number of observations $\yexp\ofi$
and in particular it is additive, i.e. the total information 
yielded by two independent experiments is the sum 
of the information from each experiment taken separately.

Second, the definition of the 
``information obtained from a set of measurements''
depends on which parameters we want to infer from them.
This is clear from Eq.~\ref{eq:infodef}, 
which defines Fisher information $\Ifx$ about $\xtruvec$
in terms of a set of derivatives 
with respect to the parameters $\xtruvec$.

\hyphenation{semi-definite}
Finally, information is related to precision:
the greater the information available
from a set of measurements about some unknown parameters,
the lower the uncertainty that can be achieved 
from the measurements on the estimation of these parameters.
More formally,
if $\xhatvec$ is any unbiased estimator of the parameter vector $\xtruvec$
derived from the measurements $\yexpvec$,
then under the same regularity conditions previously assumed it can be shown that 
$\cov(\xhatvec, \xhatvec) \succeq (\IfxMat)^{-1}$,
where the symbol $\succeq$ indicates that the difference between 
the matrices on the left and right hand sides is positive semidefinite.
In particular, for the diagonal elements of these matrices,
\begin{equation}
\label{eq:crlbdiag}
\var(\xhat\ofa) \ge (\IfxMat)^{-1}\ofaa %
\myh.
\end{equation}
In other words, the quantity $(\IfxMat)^{-1}\ofaa$
represents a lower bound (called Cramer-Rao lower bound)
on the variance of any unbiased estimator of each parameter $\xtru\ofa$.

\subsection{BLUE estimators and Fisher information}

An unbiased estimator whose variance 
is equal to its Cramer-Rao lower bound,
i.e. one for which the equality in Eq.~\ref{eq:crlbdiag} holds,
is called an efficient unbiased estimator.
While in the general case it is not always possible to build one,
an efficient unbiased estimator does exist
under the assumption that
the $\nexp$ measurements $\yexpvec$ 
are multivariate Gaussian distributed
with a positive definite covariance matrix that is known a priori
and does not depend on the unknown parameters $\xtruvec$.
This is the same assumption that had been used
for the description of the BLUE method in Ref.~\cite{bib:valassi}
and we will take it as valid throughout the rest of this paper.

As discussed at length in Refs.~\cite{bib:bos,bib:james},
such distributions possess in fact a number of special properties
that significantly simplify all statistical calculations involving them.
In particular,
it is easy to show,
in the general case with several unknown parameters,
that the best linear unbiased estimator is
under these assumptions
an unbiased efficient estimator,
i.e. that its covariance matrix is equal
to the inverse of the Fisher information matrix.
Moreover, the Fisher information matrix and the combined covariance
do not depend on the unknown parameters $\xtruvec$
under these assumptions,
while this is not true in the general case.
For Gaussian distributions,
the best linear unbiased estimator also
coincides with the maximum likelihood estimator~\cite{bib:bos},
while this is not true in most other cases,
including the case of Poisson distributions.

In the case of one unknown parameter, in particular,
i.e. when the parameter vector $\xtruvec$
reduces to a scalar $\ytru$,
the probability density function is simply
\begin{multline}
\pdf(\yexpvec;\ytru)
\, = \,
\frac{1}{(2\pi)^{\nexp/2}(\det \McovMat)^{\nexp/2}}
\\ \times \, 
\exp\bigl[ -\frac{1}{2}
\sum_{i=1}^{\nexp} \sum_{j=1}^{\nexp} 
(\yexp\ofi-\ytru) \Minv\ofij (\yexp\ofj-\ytru)
\bigr]\myh.
\end{multline}
Remembering that 
$\Mcov\ofij=\Exp[(\yexp\ofi\!-\!\ytru)(\yexp\ofj\!-\!\ytru)]$
is the covariance of the unbiased measurements $\yexp\ofi$ and $\yexp\ofj$,
the Fisher information for $\ytru$,
which also reduces to a scalar $\Ify$,
can simply be written as 
\begin{equation}
\Ify = \transpose{\Uvec}\MinvMat\Uvec 
\myh.
\label{eq:fishgauss1}
\end{equation}
This is clearly the inverse of the variance of the BLUE for $\ytru$
corresponding to the central value weights $\lambda\ofi$ given in Eq.~\ref{eq:coeff1},
\begin{equation}
  \syhat^2 =
  \sum_{i=1}^{\nexp} \sum_{j=1}^{\nexp} 
  \lambda\ofi \, \Mcov\ofij \, \lambda\ofj \, =
  \frac{\raisebox{2pt}{1}}{\raisebox{-2pt}{$\bigl(\transpose{\Uvec}\MinvMat\Uvec\bigr)$}} =
  \frac{\raisebox{2pt}{1}}{\raisebox{-2pt}{$\Ify$}}
  \myh.
\label{eq:covblue1}
\end{equation}
To further simplify the notation, in the following by $\Info$ 
we will always indicate the information $\Ify$ relative to $\ytru$, 
dropping the superscript $\ytru$.

\subsection{Intrinsic and marginal information weights}
\newcommand{\Dinfoi}{\Delta\Info_i}
\newcommand{\miw}{\mathrm{MIW}}

Having recalled the relevance of the Fisher information concept
to quantitatively assess the contribution of a set of measurements 
to the knowledge about an unknown parameter,
we may now introduce our proposal about how to best represent
the ``weight that a measurement carries in the combination''
or its ``relative importance''.
We define this in terms of intrinsic and marginal information weights.
Our approach is radically different 
from that of Refs.~\cite{bib:tev_top,bib:lhc_top}, 
because we do not attempt to make sure
that the $\nexp$ weights for the different measurements sum up to 1.

\begin{table*}[t]
\renewcommand{\arraystretch}{1.1}
\begin{center}
\vspace*{1mm}
\begin{tabular}{|lc|c|c|c|c|}
\hline
\multicolumn{2}{|c|}{Measurements} & CVW/\%  & IIW/\%  & MIW/\%  & RI/\%  \\
\hline
A &     103.00 $\pm$       3.87 &      40.00 &      40.00 &      40.00 &      40.00 \\
B &      98.00 $\pm$       3.16 &      60.00 &      60.00 &      60.00 &      60.00 \\
Correlations &      --- &      --- &      0.00 &      --- &      --- \\
\hline
BLUE / Total &     100.00 $\pm$       2.45 &     100.00 &     100.00 &     100.00 &     100.00 \\
\hline
\end{tabular}
\end{center}
\vspace*{-2mm}
\begin{center}
\begin{tabular}{|lc|c|c|c|c|}
\hline
\multicolumn{2}{|c|}{Measurements} & CVW/\%  & IIW/\%  & MIW/\%  & RI/\%  \\
\hline
A &     103.00 $\pm$       3.87 &      40.00 &      40.00 &      40.00 &      25.00 \\
B1 &      99.00 $\pm$       4.00 &      90.00 &      37.50 &      50.63 &      56.25 \\
B2 &     101.00 $\pm$       8.00 &     -30.00 &       9.38 &      22.50 &      18.75 \\
Correlations &      --- &      --- &     13.13 &      --- &      --- \\
\hline
BLUE / Total &     100.00 $\pm$       2.45 &     100.00 &     100.00 &     113.13 &     100.00 \\
\hline
\end{tabular}
\end{center}
\vspace*{-2mm}
\begin{center}
\begin{tabular}{|lc|c|c|c|c|}
\hline
\multicolumn{2}{|c|}{Measurements} & CVW/\%  & IIW/\%  & MIW/\%  & RI/\%  \\
\hline
A &     103.00 $\pm$       3.87 &      40.00 &      40.00 &      40.00 &      25.00 \\
B11 &      99.01 $\pm$       4.00 &      45.00 &      37.50 &  {\tiny $\sim$ }0 &      28.13 \\
B12 &      98.99 $\pm$       4.00 &      45.00 &      37.50 &  {\tiny $\sim$ }0 &      28.13 \\
B2 &     101.00 $\pm$       8.00 &     -30.00 &       9.37 &      22.50 &      18.75 \\
Correlations & --- & --- &     -24.37 & --- & --- \\
\hline
BLUE / Total &     100.00 $\pm$       2.45 &     100.00 &     100.00 &      62.50 &     100.00 \\
\hline
\end{tabular}
\vspace*{2mm}%
\caption{Results 
for the combination of A and B (top, $\chi^2$/ndof=1.00/1),
for that of A, B1 and B2 (center, $\chi^2$/ndof=1.17/2)
and for that of A, B11, B12 and B2 (bottom, $\chi^2$/ndof=2.42/3).
For each input measurement $i$ the following are listed: 
the central value weight CVW$_i$ or $\lambda_i$, 
the intrinsic information weight $\iiw_i$
(also shown for the correlations),
the marginal information weight $\miw_i$, 
the relative importance $\ri_i$.
In the last row in each table, 
the BLUE central value and error
and the sum of all weights in each column are displayed.}
\label{tab:iws}
\renewcommand{\arraystretch}{1}
\end{center}
\vspace*{-6mm}
\end{table*}

Formally, we define the ``intrinsic'' information weight
for each individual measurement 
simply as the ratio of the information it carries when taken alone
(the inverse of its variance)
to the total information for the set of measurements in the combination
(the inverse of the BLUE variance),
\begin{equation}
\iiw\ofi=\frac{1/\si^2}{1/\syhat^2}=\frac{1/\si^2}{I} \myh.
\end{equation}
We complement this definition by introducing a weight
carried by the ensemble of all correlations between the measurements:
\begin{equation}
\iiwcorr=\frac{1/\syhat^2-\sum_{i=1}^\nexp 1/\si^2}{1/\syhat^2}=\frac{I-\sum_{i=1}^\nexp 1/\si^2}{I} %
\label{eq:iwcorr}
\end{equation}
so that the sum of the $\nexp+1$ terms adds up to 1,
\begin{equation}
\iiwcorr + \sum_{i=1}^\nexp\iiw\ofi=1 \myh.
\end{equation}
In our opinion,
the information contribution represented by $\iiwcorr$
cannot be attributed to any of the individual measurements alone,
because it is the result of their collective interplay
through the ensemble of their correlations.
Note that we did not split this weight
into sub-contributions from one or more specific correlations because,
while this is unambiguous in some specific cases,
in general it is a complex task which implies a certain arbitrariness.

Another useful way to quantify the information 
that an individual measurement $\yexp\ofi$ brings in a combination 
is to look at its ``marginal'' information $\Dinfoi$,
i.e. the additional information available when $\yexp\ofi$ is added to a combination
that already includes the other $\nexp\!-\!1$ measurements.
We define the marginal information weight of $\yexp\ofi$ %
as the ratio of its marginal information
to the total information 
in the combination of all $\nexp$ measurements:
\begin{equation}\begin{split}
\miw_i & =
\frac{\Dinfoi}{\ I_{(\nexp\ \mathrm{meas.})}\ } 
\\ & = 
\frac{I_{(\nexp\ \mathrm{meas.})} - I_{(\nexp\!-\!1\ \mathrm{meas.\ i.e.\ all\ meas.\ except\ }i)}}{I_{(\nexp\ \mathrm{meas.})}}.
\end{split}\end{equation}
The sum of the weights for all measurements does not add up to 1,
but we do not find it appropriate
to introduce an extra weight to re-establish a normalisation condition.
The interest of a marginal information weight $\miw_i$, in fact,
is that it already accounts 
both for the information $1/\sigma_i^2$ ``intrinsically'' 
contributed by measurement $\yexp\ofi$,
and for that contributed by its correlations to all other measurements
in the presence of their correlations to one another.
The sum of the marginal information weights for all measurements %
involves a complex double-counting of these effects
and we do not find it to be a useful quantity
to easily understand the effect of correlations.

The intrinsic information weights for the different measurements
are, by construction, always positive.
The weight for the correlations
can, instead, be negative, null or positive.
In other words, according to our definition,
while every measurements
always adds intrinsic information to a combination,
the net effect of correlations may be to increase the combined error,
to keep the combined error unchanged or, less frequently, 
to decrease it.

Marginal information weights are guaranteed to be non-negative
(as discussed more in detail in Sec.~\ref{sec:inflow}),
but they are generally different 
from the corresponding intrinsic information weights
if the measurement is correlated to any of the others.
In particular, $\miw_i<\iiw_i$ represents 
the common situation where one part 
of the intrinsic information contributed by one measurement
is reduced by correlations,
while $\miw_i>\iiw_i$ represents the cases 
where its correlations 
amplify its net contribution to information.
We will discuss these issues in more detail
in Sec.~\ref{sec:simpletwomeas},
in the specific case of two measurements of one observable.

\graphicspath{{./gnuplot/}{./}}
\newcommand{\figureOne}{
\begin{figure*}[t]
\vspace*{-2mm}
\begin{center}
\mbox{%
\includegraphics[height=6.3cm]{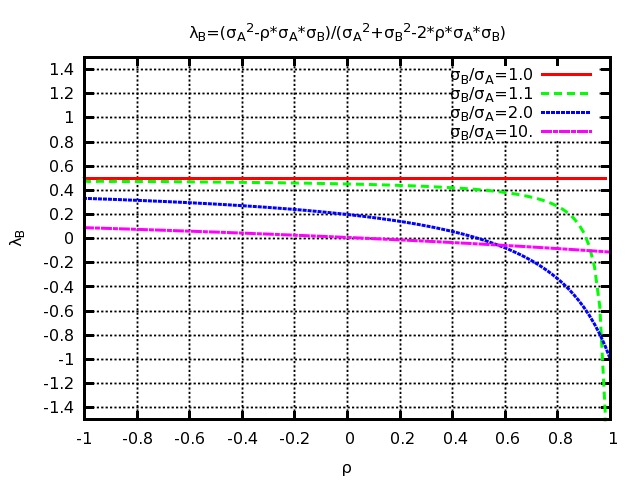} %
\includegraphics[height=6.3cm]{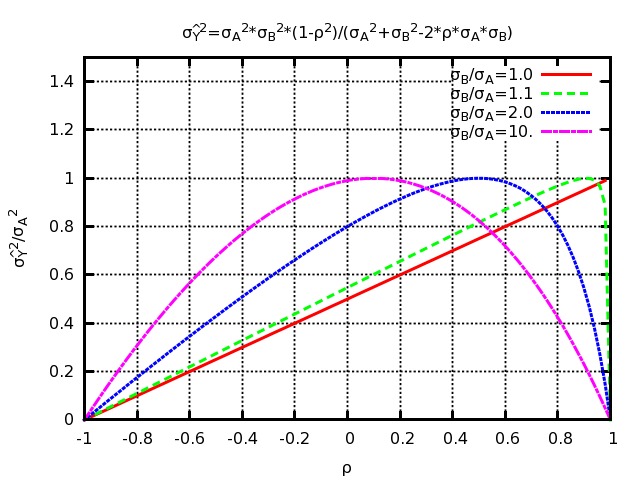} %
}
\end{center}
\vspace*{-2mm}
\caption{BLUE coefficient $\protect\lB$ for measurements B (left) 
and combined BLUE variance $\protect\syhat^2$ (right) 
as a function of the correlation $\rho$ 
between two measurements A and B 
for various fixed values of the ratio $\protect\sB/\protect\sA$.
This is essentially the same as Fig.~1 in Ref.~\cite{bib:lyons}.}
\label{fig:scVsRho}
\end{figure*}
}

In the simple example we used in Sec.~\ref{sec:ri},
the intrinsic and marginal information weights 
in the combination of the two measurements A and B 
and in the combination of the three measurements A, B1 and B2 
are summarised in the top and central sections of Table~\ref{tab:iws},
where they are compared 
to their BLUE coefficients (or central value weights)
and their ``relative importances'' according to Eq.~\ref{eq:ri}.
Note that all of these quantities 
($\iiw\ofi$, $\miw\ofi$, CVW$\ofi$, $\ri\ofi$)
coincide for both measurements in the combination of A and B,
where there are no correlations,
but they differ significantly
in the combination of A, B1 and B2,
in the presence of large positive correlations.
In particular, the intrinsic and marginal 
information weights for A are always equal to 40\%
whether A is combined to B alone or to B1 and B2 together;
conversely, the marginal information weights of B1 and B2
are significantly larger than their intrinsic information weights,
precisely because together, thanks to their correlation,
they achieve more than they could achieve individually.
Note also that the ``relative importance'' of B1
is larger than both its intrinsic and marginal information weights,
which in our opinion shows that it is clearly overestimated.

We should stress at this point that
information weights also have their own limitations
and should be used with care. In particular,
the main interest of information weights
should not be that of ranking measurements,
but rather that of providing a quantitative tool
for a better understanding of how the different measurements,
individually and together, contribute to our combined knowledge
about the parameters that we want to infer.
We believe that attempting to determine
which individual experiment provides
the ``best'' or ``most important'' contribution to a combination
is a goal of relatively limited scientific use and,
more importantly, is a question that involves
some degree of arbitrariness.
As we mentioned above, when combining $\nexp$ correlated measurements,
it is very difficult to unambiguously split $\iiwcorr$
into sub-contributions from the several correlations
that simultaneously exist between those measurements.
In particular, it would be quite complex to disentangle 
the two competing effects that each correlation
may have on the information contributed by any given measurement,
that of amplifying this contribution
through the collaboration with other measurements,
and that of reducing this contribution by making 
the measurements partially redundant with each other.
As a consequence,
``ranking'' individual measurements
by their intrinsic or marginal information weights
is a practise that we do not advocate or recommend.

\figureOne

To better illustrate what we mean,
in the bottom section of Table~\ref{tab:iws}
we have added a slightly different example, where
it is now assumed that $\yBo$ 
is itself the result of the combination 
of two very similar measurements 
$\yBoo=99.01\pm4.00$ and $\yBot=98.99\pm4.00$
that are 99.999\% correlated to each other
(and are each individually 87.5\% correlated to $\yBt$).
It is not surprising in this case that B11 and B12
have a central value weight equal to half that of B1,
an intrinsic information weight that is the same as that of B1,
but a marginal information weight that is essentially zero
(because including B12 is largely redundant %
if the almost identical measurement B11 has already been included,
and viceversa).
While in the combination of A, B1 and B2 %
the net effect of correlations was to amplify 
the information contribution of both B1 and B2
by $\miw\ofBo\!-\!\iiw\ofBo\!=\!\miw\ofBt\!-\!\iiw\ofBt\!=\!13.1\%$,
in this third example %
the information contributions of B11 and B12
are also affected by the competing effect of their mutual correlation,
which brings their $\miw$ down essentially to zero.
This example is also interesting because it clearly shows 
that very different ``rankings'' may be obtained
for the individual measurements if they are ordered
by decreasing values of $\iiw\ofi$, $\miw\ofi$, CVW$\ofi$ or $\ri\ofi$:
for instance, measurement B11 has the highest CVW and $\ri$,
the second highest $\iiw$, but the lowest $\miw$.
Excluding $\ri$, which we already argued to be an ill-defined quantity,
we see in this case that CVW, $\iiw$ and $\miw$
all have their limitations if they are used for ``ranking''.
Indeed, CVW can be negative, which may give the false impression
that a measurement makes a combination worse instead of improving it;
$\iiw$ completely ignores the effect of correlations;
$\miw$ only describes the marginal contribution 
of a single measurement and of its correlations.
For these reasons, we propose to quote all
of CVW, $\iiw$ and $\miw$ 
whenever a combination of several measurements is presented,
while explicitly refraining to use any of them 
for ranking individual measurements.

\section{Negative BLUE coefficients and ``high-correlation regimes''}
\label{sec:highcorr}

In this section, 
we use the concept of Fisher information
to explore the relation between negative BLUE coefficients
and the size of correlations between measurements.
We start by revisiting the discussion of these issues
presented in Ref.~\cite{bib:lyons} for two measurements of one parameter,
whose conclusion was that negative weights appear
when the positive correlation between the two measurements 
exceeds a well-defined threshold.
We then generalize this conclusion to $\nexp$ measurements
of one parameter, first by computing marginal information 
and then by analysing the derivatives of Fisher information
with respect to the correlations between measurements:
we show, in particular, that negative central value weights
in BLUE combinations are always a sign %
of a ``high-correlation'' regime, 
where the marginal effect of further increasing one or more of these correlations
is that of reducing the errors on the combined estimates 
rather than increasing them. In Sec.~\ref{sec:conservativeness} we will
discuss important practical consequences of what is presented
in this section.

\subsection{The simple case of two measurements of one parameter}
\label{sec:simpletwomeas}

In the simple case of two measurements A and B
of a single physical quantity $\ytru$,
the coefficients in the BLUE weighted average 
$\yhat\!=\!\lA\yA+\lB\yB$ are simply given by
\begin{eqnarray}
  \lA & = & \frac{\sB^2-\rho\sA\sB}{\sA^2+\sB^2-2\rho\sA\sB} \myh, \\
  \lB & = & \frac{\sA^2-\rho\sA\sB}{\sA^2+\sB^2-2\rho\sA\sB} \myh, 
\label{eq:l2}
\end{eqnarray}
and the combined variance $\syhat^2$
(i.e. the inverse of the Fisher information) by
\begin{equation}
  \syhat^2 = \frac{\sA^2\sB^2(1-\rho^2)}{\sA^2+\sB^2-2\rho\sA\sB} 
  = \frac{1}{\Info} \myh,
\label{eq:syhat2}
\end{equation}
where $\sA$ and $\sB$ are the errors on the two measurements
and $\rho$ is their correlation.
Assuming that the two errors are fixed, with $\sB>\sA$,
the functional dependency on the correlation $\rho$
of the BLUE coefficient $\lB$ for measurement B
and of the combined variance $\syhat^2$,
given by Eq.~\ref{eq:l2} and Eq.~\ref{eq:syhat2} respectively,
are shown in Fig.\ref{fig:scVsRho}, 
left and right respectively,
for various values of the ratio $\sB/\sA$.
For positive correlations, as discussed in Ref.~\cite{bib:lyons},
the combined BLUE variance increases 
from $\syhat^2\!=\!1/(1/\sA^2+1/\sB^2)$ at $\rho\!=\!0$
to a maximum value of $\syhat^2\!=\!\sA^2$ at $\rho\!=\!\sA/\sB$
and it then decreases to $\syhat^2\!=\!0$ in the limit of $\rho\!=\!1$.
Therefore, for combinations
with large correlations among measurements,
the combined uncertainty strongly depends on $\rho$, and is expected 
to vanish at $\rho\!=\!1$; this
implies that, close to these regions, determining the correlation
with high accuracy is mandatory so as not to bias the combination.
The BLUE coefficient for B steadily decreases 
from $\lB\!=\!(1/\sB^2)/(1/\sA^2+1/\sB^2)$ at $\rho\!=\!0$
to a negative value of $\lB\!=\!-1/(\sB/\sA-1)$ in the limit of $\rho\!=\!1$,
passing through $\lB\!=\!0$ at the correlation 
$\rho\!=\!\sA/\sB$ where $\syhat^2$ is maximized.

\newcommand{\figureTwo}{
\begin{figure*}[htb]
\vspace*{-2mm}
\begin{center}
\mbox{%
\includegraphics[height=6.3cm]{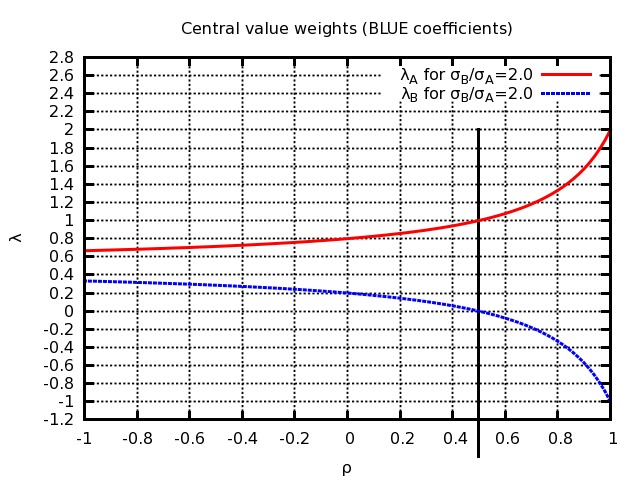} %
\includegraphics[height=6.3cm]{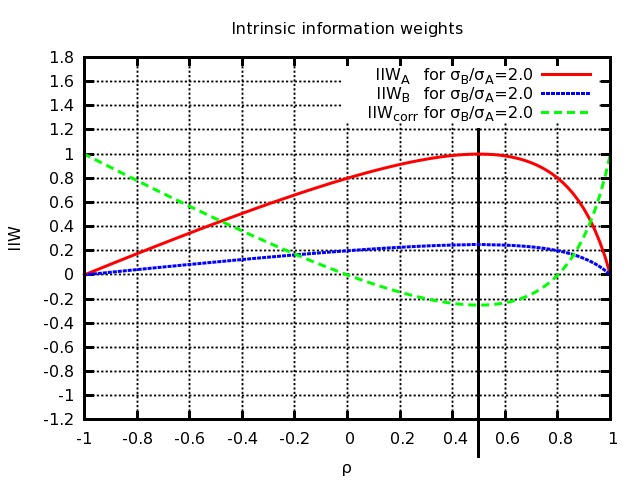} %
}
\end{center}
\vspace*{-13mm}
\begin{center}
\mbox{\hspace*{-6mm}
\includegraphics[height=6.3cm]{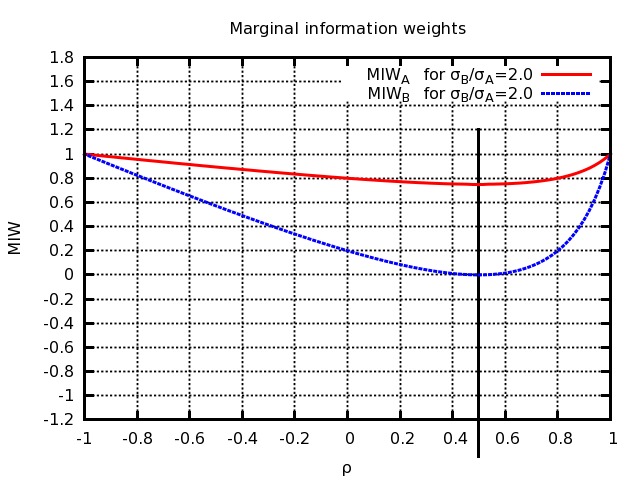} %
\includegraphics[height=6.3cm]{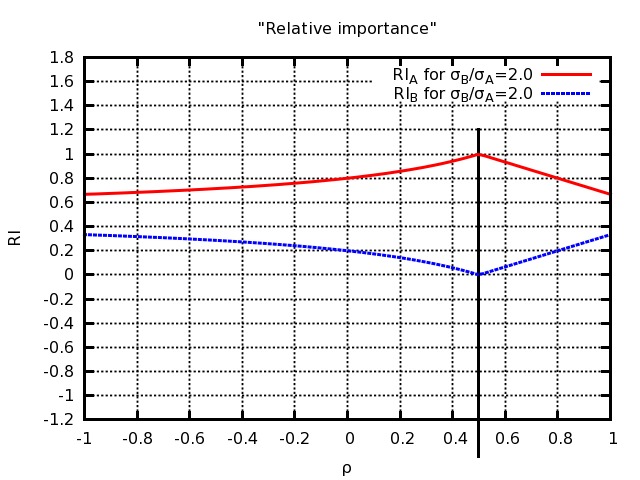} %
}
\end{center}
\vspace*{-2mm}
\caption{BLUE coefficients $\lambda$ for A and B (top left),
intrinsic information weights $\iiw$ for A, B and correlations (top right),
marginal information weights $\miw$ for A and B (bottom left)
and relative importances $\ri$ for A and B (bottom right)
as a function of the correlation $\rho$ between A and B
for the specific example $\protect\sB/\protect\sA$=2.
The black line in each plot indicates the boundary
between low-correlation and high-correlation regimes
at $\rho\!=\!1/2$ where the total information from A and B is minimized.}
\label{fig:WeightsVsRho}
\end{figure*}
}

In other words,
the threshold value $\rho\!=\!\sA/\sB$
effectively represents a boundary between two regimes,
a ``low-correlation regime'',
where $\lB$ is positive and $\syhat^2$ increases as $\rho$ grows,
and a ``high-correlation regime'',
where $\lB$ is negative and $\syhat^2$ decreases as $\rho$ grows.
Note that the BLUE variance
from the combination of A and B at the boundary 
between the two regimes $\rho\!=\!\sA/\sB$
is equal to that from A alone ($\syhat^2\!=\!\sA^2$), 
while it is lower on either side of the boundary.
In the same way, 
the Fisher information from the combination
at the boundary between the two regimes
is equal to that from A alone, 
while it is higher on either side of the boundary:
in other words, 
the marginal contribution to information 
from the addition of B in the combination
is zero at the boundary, 
but it is positive on either side of it.
Note in passing that the fact that 
the BLUE coefficient for B is zero
does not mean however that the measurement
is simply not used in the combination,
because the central value measured by B
does in any case contribute to the calculation
of the overall $\chi^2$ for the combination:
this statement remains valid 
for the combination of $\nexp$ measurements,
although we will not repeat it in the following.

A possible interpretation of the transition at $\rho\!=\!\sA/\sB$,
which will become useful later on and
is complementary to that given in Ref.~\cite{bib:lyons}
(as well as to that given in Ref.~\cite{bib:cox}
using the Cholesky decomposition formalism),
is the following.
In the low-correlation regime $\rho\leq\sA/\sB$,
the full covariance matrix can be written as
the sum of two positive-definite components: one 
that is common to A and B, i.e. 100\% correlated and 
with the same size in both, and one that is uncorrelated:
\begin{multline} %
\renewcommand{\arraystretch}{1.2}
\biggl(\begin{array}{cc}
\sA^2 & \,\rho\sA\sB\, \\ \,\rho\sA\sB\, & \sB^2 \\
\end{array}\biggr)
=
\biggl(\begin{array}{cc}
\,\rho\sA\sB\, & \,\rho\sA\sB\, \\ \,\rho\sA\sB\, & \,\rho\sA\sB\, \\
\end{array}\biggr)_\mathrm{com}
\\ +
\biggl(\begin{array}{cc}
\,\sA^2\!-\!\rho\sA\sB\, & 0 \\ 0 & \,\sB^2\!-\!\rho\sA\sB\, \\
\end{array}\biggr)_\mathrm{unc}
\myh.
\renewcommand{\arraystretch}{1}
\label{eq:split1}
\end{multline} %
Indeed, only when the off-diagonal covariance $\rho\sA\sB$ is smaller
than both variances $\sA^2$ and $\sB^2$
can the ``uncorrelated'' error component be positive definite.
The possibility to split the covariance matrix in this way
can be interpreted by saying that, in the low-correlation regime,
the marginal information 
added to the combination by the less precise measurement B
comes from its contribution of 
independent (uncorrelated) knowledge about the unknown parameter.
To combine A and B in this case, in fact, 
one may simply think of temporarely ignoring
the irreducible common component of the error,
combining the two measurements based only 
on the uncorrelated error components
and finally adding back the common error component:
it is easy to see that this would lead 
to a total combined variance
\begin{multline} %
\renewcommand{\arraystretch}{1.2}
  \syhat^2 = \biggl(\rho\sA\sB\biggr)_\mathrm{com} \\ 
  + \biggl(\frac{\sA\sB(\sA\!-\!\rho\sB)(\sB\!-\!\rho\sA)}
  {\sA^2+\sB^2-2\rho\sA\sB}\biggr)_\mathrm{unc},
\renewcommand{\arraystretch}{1}
\label{eq:syhat2lc}
\end{multline}
which adds up to the same value given in Eq.~\ref{eq:syhat2}.
With respect to the measurement of A taken alone,
adding B helps in this case 
by reducing the uncorrelated error component
(the second term on the right-hand side).

\figureTwo

In the high correlation regime, conversely,
the covariance matrix can not be seen 
as the sum of a common error and an uncorrelated error
as in Eq.~\ref{eq:split1}.
Instead, the full covariance matrix can be written as
the sum of a component common to A and B,
i.e. 100\% correlated and with the same size in both,
and of another systematic effect that is also 
100\% correlated, but has different sizes in A and B:
\begin{multline} %
\renewcommand{\arraystretch}{1.2}
\biggl(\begin{array}{cc}
\sA^2 & \,\rho\sA\sB\, \\ \,\rho\sA\sB\, & \sB^2 \\
\end{array}\biggr)
=
\biggl(\begin{array}{cc}
\syhat^2 & \syhat^2 \\ \syhat^2 & \syhat^2 \\
\end{array}\biggr)_\mathrm{com}
\\ +
(\sA^2+\sB^2-2\rho\sA\sB)
\biggl(\begin{array}{cc}
\lB^2 & -\lA\lB \\ -\lA\lB & \lA^2 \\
\end{array}\biggr)_\mathrm{cor}
\myh.
\renewcommand{\arraystretch}{1}
\label{eq:hcsplit}
\end{multline} %
In the combined result,
the total error $\syhat$ comes exclusively
from the common systematic uncertainty in the first component,
while the contribution from the correlated systematic uncertainty 
in the second component is 0.
In other words,
\begin{equation}
  \syhat^2 = \biggl(\syhat^2\bigg)_\mathrm{com} + \biggl( 0 \biggr)_\mathrm{cor},
\label{eq:syhat2hc}
\end{equation}
which can again be seen by removing the common component,
combining and then adding it back at the end.
For all practical purposes,
one can thus say that,
in the high correlation regime,
the marginal information added by the less precise measurement B
does not come from independent knowledge it contributes 
about the unknown parameter, but %
from its ability to constrain and remove 
a systematic uncertainty that also affects A,
but to which B has a larger sensitivity.
With respect to the measurement of A taken alone,
in fact, adding B helps
by completely getting rid 
of the correlated error component
(the second term on the right-hand side).
Note also, as discussed in Refs.~\cite{bib:lyons,bib:cox},
that the two individual measurements A and B
are on the same side of the combined estimate
in the high correlation regime (unless they coincide),
because $\lB\!<\!0$ implies that
$\yhat\!<\!\yA\!<\!\yB$ or $\yB\!<\!\yA\!<\!\yhat$.

To further illustrate this difference
between the low and high correlation regimes,
it is interesting to study 
the functional dependency on the correlation $\rho$
of the intrinsic and marginal information weights $\iiw$ and $\miw$
that we introduced in Sec.~\ref{sec:information}.
In Fig.~\ref{fig:WeightsVsRho}
we compare the functional dependencies on $\rho$
of the BLUE coefficients $\lambda$,
intrinsic information weights $\iiw$,
marginal information weights $\miw$
and relative importances $\ri$
for one specific example where $\sB/\sA\!=\!2$.
As expected, both the intrinsic and marginal information weights 
of the two individual measurements are non-negative
and actually coincide with each other 
and with the BLUE coefficients when $\rho\!=\!0$,
while they have two maxima and two minima, respectively, 
at the boundary between low-correlation and high-correlation regimes
at $\rho\!=\!\sigma_A/\sigma_B$ where 
the total information from A and B is minimized.
In the limit of extremely high correlation $\rho\!=\!1$,
where the combined variance tends to 0
as the information contributed by the correlation
between the two measurements tends to infinity,
the intrinsic information weights tend to 0
while the marginal information weights tend to 1,
because the intrinsic information contributed
by each experiment individually is negligible
with respect to the large contribution
they achieve together through their correlation.
For comparison, note instead that the ``relative importances''
of A and B at $\rho\!=\!1$ are both positive and sum up to 1
while being different from each other,
which in our opinion is another indication that
this concept fails to acknowledge the relevance here
of the information contribution from correlations.

\newcommand{\Rcov}{\mathcal{D}}
\newcommand{\Rinv}{\mathcal{D}^{-1}}
\newcommand{\RcovMat}{\mymat{\mathcal{D}}}
\newcommand{\RinvMat}{\mymat{\mathcal{D}}^{-1}}
\newcommand{\rcov}{d}
\newcommand{\cvec}{\myvec{c}}
\newcommand{\uvec}{\myvec{u}}
\newcommand{\myparl}{(}
\newcommand{\myparr}{)}
\newcommand{\ith}{i^{\mathrm{th}}}
\newcommand{\jth}{j^{\mathrm{th}}}
\newcommand{\nth}{\nexp^{\mathrm{th}}}

\subsection{Marginal information from the $\ith$ measurement of one parameter}
\label{sec:inflow}

\begin{sloppypar}
To generalize the concepts of low and high correlation regimes,
and show their relation to negative BLUE coefficients,
in the more general case of $\nexp$ measurements of one parameter,
we now derive formulas
to calculate the marginal information $\Dinfoi$
of the $\ith$ measurement in an $\nexp$-measurement combination,
using the ``information inflow'' formalism of Ref.~\cite{bib:bos}.
Without loss of generality,
imagine that the $\nexp$ measurements in the combination are reordered 
so that the $\ith$ measurement we are interested in
becomes the last, i.e. the $\nth$, measurement.
The full covariance matrix for all $\nexp$ measurements 
can then be written as
\end{sloppypar}
\begin{equation}
\McovMat = 
\left(
\begin{array}{cc}
\RcovMat & \cvec \\
\transpose{\cvec} & \rcov \\
\end{array}
\right)
\myh,
\label{eq:Ddc}
\end{equation}
where the variance of the $\ith$ measurement 
is given by $\sigma_i^2\!=\!\rcov$,
its covariances with all other measurements 
are the $(\nexp\!-\!1)$ elements of the vector $\cvec$,
and $\RcovMat$ is the $(\nexp\!-\!1)\!\times\!(\nexp\!-\!1)$ 
covariance matrix of these $\nexp\!-\!1$ other measurements.
Using Frobenius' formula~\cite{bib:bos}, 
the inverse of this matrix can be written as
\begin{equation}
\renewcommand{\arraystretch}{1.5}
\MinvMat = 
\left(
\begin{array}{cc}
\hspace*{0.2cm} 
\RinvMat + 
\frac{\left(\RinvMat\cvec\right)\left(\transpose{\cvec}\RinvMat\right)}
{\raisebox{-2pt}{$\rcov-\myparl\transpose{\cvec}\RinvMat\cvec\myparr$}} 
\hspace*{0.2cm} & \hspace*{0.2cm} 
\frac{-\left(\RinvMat\cvec\right)}
{\raisebox{-2pt}{$\rcov-\myparl\transpose{\cvec}\RinvMat\cvec\myparr$}}
\hspace*{0.2cm} \\ \hspace*{0.2cm}
\frac{-\left(\transpose{\cvec}\RinvMat\right)}
{\raisebox{-2pt}{$\rcov-\myparl\transpose{\cvec}\RinvMat\cvec\myparr$}}
\hspace*{0.2cm} & \hspace*{0.2cm} 
\frac{1}
{\raisebox{-2pt}{$\rcov-\myparl\transpose{\cvec}\RinvMat\cvec\myparr$}}
\hspace*{0.2cm} \\
\end{array}
\right)
\myh,
\label{eq:DdcInv}
\renewcommand{\arraystretch}{1}
\end{equation}
where $1/(\rcov-\myparl\transpose{\cvec}\RinvMat\cvec\myparr) > 0$
because it is a diagonal element of the inverse 
of the symmetric positive definite matrix $\McovMat$.
Keeping in mind that the information $\Info$
from the combination of $\nexp$ measurements
according to Eq.~\ref{eq:fishgauss1}
is simply $\transpose{\Uvec}\MinvMat\Uvec$,
it can easily be shown~\cite{bib:bos}
that the marginal information $\Dinfoi$ 
(or information inflow) from the $\ith$ measurement
is given by
\begin{equation}
\Dinfoi =
\frac{\left(\myparl\transpose{\uvec}\RinvMat\cvec\myparr-1\right)^2}
{\rcov-\myparl\transpose{\cvec}\RinvMat\cvec\myparr}
\ge 0
\myh,
\label{eq:inflow}
\end{equation}
where $\uvec$ is a vector whose $(\nexp\!-\!1)$ elements are all equal to 1
(i.e. the equivalent of $\Uvec$, but with one less measurement).

\begin{sloppypar}
Let us now analyze this formula, assuming that 
the covariance matrix $\RcovMat$ of the other $(\nexp\!-\!1)$ measurements 
and the variance $\rcov$ of the $\ith$ measurement are both fixed,
while the $(\nexp\!-\!1)$ correlations in $\cvec$ can vary.
We then observe that the condition $\Dinfoi\!=\!0$,
where information has a minimum, corresponds to a hyperplane 
in the $(\nexp\!-\!1)$-dimensional space of these correlations, 
defined by
\begin{equation}
\Dinfoi = 0 \iff
\myparl\transpose{\uvec}\RinvMat\cvec\myparr-1 = 0
\myh.
\label{eq:hyperplane}
\end{equation}
This hyperplane divides the $(\nexp\!-\!1)$-dimensional space of correlations
into two half-spaces:
a half-space containing the origin $\cvec\!=\!\myvec{0}$
(i.e. the point where there are no correlations),
which we will therefore call the ``low-correlation'' regime;
and a half-space that does not contain the origin,
which we will call the ``high-correlation'' regime,
defined by
\begin{equation}
\myparl\transpose{\uvec}\RinvMat\cvec\myparr-1 \ge 0
\myh.
\label{eq:halfspacehighcorr}
\end{equation}
Keeping in mind that the BLUE coefficients are given by Eq.~\ref{eq:coeff1},
i.e. by
\begin{equation}
  \lvect = \frac{\transpose{\Uvec}\MinvMat}{\Info}
  \myh,
  \label{eq:coeff1bis}
\end{equation}
and substituting $\MinvMat$ by the expression in Eq.~\ref{eq:DdcInv},
we observe that the BLUE coefficient for the $\ith$ measurement is
\begin{equation}
\lambda_i =
-\frac{1}{\Info}\times
\frac{\myparl\transpose{\uvec}\RinvMat\cvec\myparr-1}
{\rcov-\myparl\transpose{\cvec}\RinvMat\cvec\myparr}
\myh.
\label{eq:lambdan}
\end{equation}
Having already observed 
that $(\rcov-\myparl\transpose{\cvec}\RinvMat\cvec\myparr)>0$,
this implies that 
\begin{equation}
\lambda_i \le 0 \iff
\myparl\transpose{\uvec}\RinvMat\cvec\myparr-1 \ge 0
\myh.
\label{eq:halfspace}
\end{equation}
Comparing this to Eq.~\ref{eq:halfspacehighcorr},
this shows that the condition $\lambda_i \le 0$
coincides with that defining the half-space
corresponding to the ``high-correlation'' regime.
In other words, the BLUE coefficient for the $\ith$ measurement
is negative if and only if its correlations to the other
$\nexp\!-\!1$ measurements are higher than the thresholds
for crossing over into the high-correlation regime;
it is instead zero on the boundary between the two regimes,
i.e. if and only if these correlations are such that the measurement
contributes no additional information to the combination.
\end{sloppypar}

The results presented above are interesting not only to point out the relation 
between negative BLUE coefficients and high-correlation regimes,
but also because they provide a formula 
for computing marginal information weights.
For the $\ith$ measurement, this is simply equal to
\begin{equation}
\miw_i = \frac{\Dinfoi}{I} = 
\frac{1}{I}\times
\frac{\left(\myparl\transpose{\uvec}\RinvMat\cvec\myparr-1\right)^2}
{\rcov-\myparl\transpose{\cvec}\RinvMat\cvec\myparr}
= \lambda_i^2 I \left(\rcov-\myparl\transpose{\cvec}\RinvMat\cvec\myparr\right) \myh.
\end{equation}
Keeping in mind that the intrinsic information weight 
for the same measurement is $\iiw_i=(1/\sigma_i^2)/I=1/dI$,
this implies that
\begin{equation}
\miw_i \ \times \ \iiw_i
= \lambda_i^2 \left(1 - \frac{\myparl\transpose{\cvec}\RinvMat\cvec\myparr}{\rcov}\right)
\le \lambda_i^2 \myh,
\end{equation}
which provides an interesting relationship
between intrinsic information weights,
marginal information weights
and central value weights.
Note, in particular,
that the equality sign holds
if the $i^{\mathrm{th}}$ measurement is not correlated to any other measurements
(i.e. if $\cvec$ is the null vector),
in which case all three weights coincide as seen in Table~\ref{tab:iws}.

\newcommand{\Amat}{\mymat{\mathcal{A}}}
\newcommand{\Ainv}{\mymat{\mathcal{A}}^{-1}}
\newcommand{\vvar}{v}
\newcommand{\vvarvec}{\myvec{\vvar}}
\newcommand{\ofp}{_p}
\newcommand{\ofq}{_q}
\newcommand{\ofl}{_{l}}
\newcommand{\ofil}{_{il}}
\newcommand{\ofjl}{_{jl}}
\newcommand{\ofkl}{_{kl}}

\subsection{Information derivatives}
\label{sec:derivatives}

In the first part of this section, we have described 
the boundary between low-correlation and high-correlation regimes
in the simplest case of the combination of two measurements,
as well as in the more complex but still specific case
of the combination of $\nexp$ measurements,
where only the $\nexp\!-\!1$ correlations of the $\ith$ measurement 
to all of the others are allowed to vary.
We now analyze the most general case
of the combination of $\nexp$ measurements of one parameter,
as a function of the $\nexp(\nexp\!-\!1)/2$
correlations of all the measurements to one another.
We do this by studying the first derivatives of information
with respect to these correlations $\rho\ofij$.

Let us consider the linear dependency of the covariance matrix $\McovMat$ 
on the $\nexp(\nexp\!-\!1)/2$ correlations $\rho\ofij$ 
between any two distinct measurements $\yexp\ofi$ and $\yexp\ofj$,
assuming instead that the variances $\Mcov\ofii$ are fixed.
Applying the generic formula~\cite{bib:bos} 
for the first derivatives of the inverse of a non-singular square matrix 
with respect to the elements of a vector it depends on, we find that
\begin{equation}
\frac{\partial\MinvMat}{\partial\rho\ofij} =
-\MinvMat\frac{\partial\McovMat}{\partial\rho\ofij}\MinvMat \myh.
\label{eq:d1A}
\end{equation}
Keeping in mind that $\Info=\transpose{\Uvec}\MinvMat\Uvec$
and that $\transpose{\Uvec}\MinvMat=\Info\lvect$
according to Eqs.~\ref{eq:fishgauss1} and~\ref{eq:coeff1bis}, respectively,
the derivatives of information with respect 
to the correlations $\rho\ofij$ can be written as
\begin{equation}
\frac{\partial\Info}{\partial\rho\ofij}
= -\transpose{\Uvec}\MinvMat\frac{\partial\McovMat}{\partial\rho\ofij}\MinvMat\Uvec
= -\Info^2\lvect\frac{\partial\McovMat}{\partial\rho\ofij}\lvec 
\myh.
\label{eq:d1Ibis}
\end{equation}
Under our assumption that only the off-diagonal covariances 
$\Mcov\ofij$ may vary while the variances $\Mcov\ofii$ are fixed,
the derivatives of the covariance matrix $\McovMat$ 
with respect to the correlations $\rho\ofij$ are
\begin{equation}\begin{split}
\left(\frac{\partial\McovMat}{\partial\rho\ofij}\right)\ofkl 
& =
\sqrt{\Mcov\ofii}\sqrt{\Mcov\ofjj} 
\times \left( \delta\ofil\delta\ofjk + \delta\ofik\delta\ofjl \right) 
\\ & = 
\sqrt{\Mcov\ofii}\sqrt{\Mcov\ofjj} 
\times \left\{
\begin{array}{ll}       
1 & \mathrm{if}\,i=k\,\mathrm{and}\,j=l\,,\\
1 & \mathrm{if}\,i=l\,\mathrm{and}\,j=k\,,\\
0 & \mathrm{otherwise}\,.
\end{array}
\right.
\label{eq:d1M}
\end{split}\end{equation}
The derivatives of information with respect 
to the correlations $\rho\ofij$ 
are then simply given by
\begin{equation}
\frac{\partial\Info}{\partial\rho\ofij} =
-2\Info^2\lambda\ofi\lambda\ofj\sqrt{\Mcov\ofii}\sqrt{\Mcov\ofjj} \myh ,
\label{eq:d1Iter}
\end{equation}
where the factor 2 comes from the fact that the covariance matrix is symmetric
and has twice as many off-diagonal elements
as there are independent correlations.

Equation~\ref{eq:d1Iter} clearly shows that, if all BLUE coefficients are positive,
the first derivatives of information are always negative with respect to 
the correlations between any two measurements,
i.e. information can only decrease if correlations are further increased:
this is the equivalent in $\nexp(\nexp\!-\!1)/2$ dimensional space
of what we have previously called a ``low-correlation'' regime,
as this sub-space is guaranteed to contain
the point where all correlations are zero.
Conversely, if at least one BLUE coefficient is negative
(and keeping in mind that they can not be all negative),
then at least one information derivative must be positive,
i.e. there is at least one correlation 
which leads to higher information if it is increased:
this is the equivalent %
of what we have previously called a ``high-correlation'' regime.
The boundary between the two regimes is a hypersurface
in $\nexp(\nexp\!-\!1)/2$ dimensional space,
defined by the condition 
that at least one BLUE coefficient is zero,
while all the others are non-negative:
when this condition is satisfied,
the information derivatives with respect
to one or more correlations are also zero,
meaning that information has reached a minimum
in its partial functional dependency on those correlations.
This completes the generalization to several measurements of one observable
of the discussion presented in Ref.~\cite{bib:lyons} for only two measurements.
Note finally that in that case, i.e. for $\nexp\!=\!2$,
all these considerations are trivially illustrated
by Figures~\ref{fig:scVsRho} and~\ref{fig:WeightsVsRho},
showing that the boundary between low and high correlation regimes
in the 1-dimensional space of the correlation $\rho$ 
is a 0-dimensional hypersurface (a point) at the value $\rho=\sA/\sB$.

\newcommand{\sunc}{\sigma_{\mathrm{(unc)}}}
\newcommand{\scor}{\sigma_{\mathrm{(cor)}}}
\newcommand{\sAunc}{\sigma_{A (\mathrm{unc})}}
\newcommand{\sAcor}{\sigma_{A (\mathrm{cor})}}
\newcommand{\sBunc}{\sigma_{B (\mathrm{unc})}}
\newcommand{\sBcor}{\sigma_{B (\mathrm{cor})}}
\newcommand{\rcor}{\rho_{\mathrm{cor}}}

\section{``Conservative'' estimates of correlations in BLUE combinations}
\label{sec:conservativeness}

A precise assessment 
of the correlations that need to be used
as input to BLUE combinations is often very hard.
Ideally, one should aim to measure these correlations
in the data or by using Monte Carlo methods.
This, however, turns out to be often impractical, if not impossible,
for instance when combining results 
produced by different experiments that use
different conventions for assessing the systematic
errors on their measurements, or when trying to combine
results from recent experiments to older results for
which not enough details were published and the expertise 
and the infrastructure to analyse the data are no longer available. 
In these situations, it may be unavoidable
to combine results using input covariance matrices
where the correlations 
between the different measurements
have only been approximately estimated,
rather than accurately measured.
In the following, we will refer to these estimates 
of correlations as the ``nominal'' correlations
(and we will extensively study the effect on BLUE combination results
of reducing correlations below these initial ``nominal'' values).

In particular, 
it is not uncommon to read in the literature
that correlations have been ``conservatively'' assumed to be 100\%.
In this section, we question the validity of this kind of statement.
A ``conservative'' estimate of a measurement error should
mean that, in the absence of more precise assessments, an overestimate 
of the true error (at the price of losing some 
of the available information from a measurement)
is more acceptable than taking the risk of claiming 
that a measurement is more accurate than it really is. 
Likewise, by ``conservative'' estimate of a correlation,
one should mean an estimate which is more likely 
to result in an overall larger combined error
than in a wrong claim of smaller combined errors.

When BLUE coefficients are all positive,
i.e. in a low-correlation regime,
information derivatives are negative and
the net effect of increasing any correlation
can only be that of reducing information
and increasing the combined error: in this case, 
choosing the largest possible positive correlations (100\%)
is clearly the most conservative choice.
Our discussion in the previous section, however,
shows that negative BLUE coefficients are a sign %
of a high-correlation regime,
where the net effect of increasing some of these correlations
is that of increasing information
and reducing the combined error:
in other words, if correlations are estimated as 100\% 
and negative BLUE coefficients are observed,
it is wrong to claim that 
correlations have been estimated ``conservatively''.

\newcommand{\postponeZero}{
In this section, 
we will first analyse under which conditions
it is indeed conservative to assume that correlations are 100\%,
using a simple two-measurement combination as an example.
For those situations where a precise evaluation of correlations is impossible,
and where setting them to their ``nominal'' estimates
would result in negative BLUE coefficients,
we will then offer a few guidelines and tools to help physicists 
make more conservative estimates of correlations.
}

\begin{figure*}[t]
\vspace*{-2mm}
\begin{center}
\includegraphics[height=6.3cm]{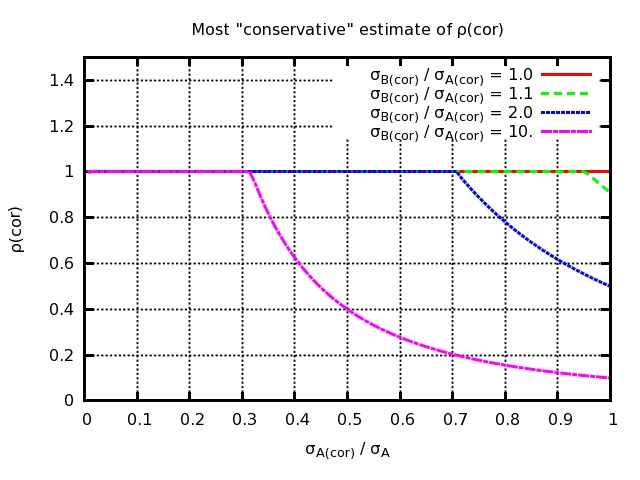} %
\end{center}
\vspace*{-4mm}
\caption{Most ``conservative'' value of an unknown correlation $\rcor$
between $\sAcor$ and $\sBcor$ as a function of $\sAcor/\protect\sA$,
for several values of $\protect\sBcor/\protect\sAcor\geq1$.}
\vspace*{-2mm}
\label{fig:rhoCons}
\end{figure*}

\postponeZero

\subsection{Conservative estimates of correlations in a two-measurement combination}
\label{sec:rhocons}

\begin{sloppypar}
Let us consider the combination of two measurements A and B, 
whose errors are well known,
but where the correlation $\rho$ between them 
could not be precisely determined.
We want to determine in this case which would be
the most ``conservative'' estimate for $\rho$. %
We do this by studying the functional dependency 
of the total combined error on $\rho$
(which throughout this Sec.~\ref{sec:rhocons} is taken as a variable parameter,
rather than any ``nominal'' estimate of the correlation). %
We observed in Sec.~\ref{sec:simpletwomeas} that 
this combination remains in a low-correlation regime
as long as the off-diagonal covariance $\rho\sA\sB$ 
is smaller than both variances $\sA^2$ and $\sB^2$,
\begin{equation}
\renewcommand{\arraystretch}{1.2}
\biggl\{\begin{array}{ccc}
\rho\sA\sB & \leq & \sA^2 \myh, \\
\rho\sA\sB & \leq & \sB^2 \myh. \\
\end{array}\biggr.
\renewcommand{\arraystretch}{1}
\label{eq:rhocons2}
\end{equation}
Let us now assume that there are only two sources of uncertainty, 
an uncorrelated error $\sunc$, e.g. of statistical origin, 
and a single systematic effect $\scor$ whose correlation 
between the two measurements is $\rcor$.
The most conservative estimate for $\rcor$
according to Eq.~\ref{eq:rhocons2} above 
is thus the largest value of $\rcor$ such that
\begin{equation}
\renewcommand{\arraystretch}{1.2}
\biggl\{\begin{array}{ccc}
\rho\sA\sB = \rcor\sAcor\sBcor & \leq & (\sAunc^2+\sAcor^2)=\sA^2 \myh,\\
\rho\sA\sB = \rcor\sAcor\sBcor & \leq & (\sBunc^2+\sBcor^2)=\sB^2 \myh.\\
\end{array}\biggr.
\renewcommand{\arraystretch}{1}
\end{equation}
This expression is very interesting because 
it is automatically satisfied by any value of $\rcor$
for measurements that are statistically dominated, 
i.e. where each of $\sAunc$ and $\sBunc$
is much larger than both $\sAcor$ and $\sBcor$:
in other words, for statistically dominated measurements, 
it is indeed a correct statement to say 
that ``correlations are conservatively assumed to be 100\%''.
\end{sloppypar}

If the measurements are not statistically dominated, 
however, the situation is different.
Taking $\sAcor$  to be the smaller 
of the two correlated errors, i.e. $\sAcor\leq\sBcor$, 
then the second condition is automatically true,
while the first condition is satisfied if and only if
\begin{equation}
\rcor\leq\frac{\sA^2}{\sAcor\sBcor}=
\frac{\sAcor/\sBcor}{(\sAcor/\sA)^2} \myh.
\end{equation}
This shows that,
when systematic errors cannot be ignored 
and the sensitivities of A and B to the correlated systematic effect 
are so different that $(\sBcor/\sAcor)\geq(\sA/\sAcor)^2$,
then it is no longer correct to take 100\%
as the most conservative value of $\rcor$,
and one must choose a correlation that is smaller than 100\%.
This is shown in Fig.~\ref{fig:rhoCons},
where the most conservative value of $\rcor$ 
is plotted as a function of $\sAcor/\sA$,
for several values of $\sBcor/\sAcor$.

This figure shows that there are two different regimes.
When $(\sBcor/\sAcor)<(\sA/\sAcor)^2$,
the most conservative value of $\rcor$ is 1:
in this case,
both measurements A and B contribute 
to the combination with positive BLUE coefficients
because, no matter how large $\rcor$ is,
the combination always remains in a low-correlation regime.
When $(\sBcor/\sAcor)\geq(\sA/\sAcor)^2$, instead,
the most conservative value of $\rcor$ is smaller than 1:
in this case, %
the combination is at the boundary
of low and high correlation regimes,
where the combined error is maximised and equal to $\syhat\!=\!\sA$
while $\lambda_A\!=\!1$ and $\lambda_B\!=\!0$,
which is more or less equivalent
(modulo the effect on $\chi^2$ previously discussed)
to excluding the less precise measurement B from the combination.

\newcommand{\sfs}{f}
\subsection{Identifying the least conservative correlations between $\nexp$ measurements}
\label{sec:id}

While it is relatively straightforward to determine 
the ``most conservative'' estimate of correlations
with only two measurements,
things get more complicated when the combination includes $\nexp$ measurements,
as the number of inter-measurement correlations 
increases to $\nexp(\nexp\!-\!1)/2$
and the ``most conservative'' value for any of them %
would depend on the values of the others.
Instead of treating all correlations as completely free parameters
as we did in Sec.~\ref{sec:rhocons},
throughout this Sec.~\ref{sec:id}
we will then suggest ways to analyse how ``conservative''
an existing ``nominal'' estimate of correlations is.
In particular, we will consider the most general case where
the full ``nominal'' covariance matrix
is built as the sum of $\nerr$ sources of uncertainty,
each with a different set 
of ``nominal'' correlations between the measurements,
\begin{equation}
\Mcov\ofij = \sum_{\ierr=1}^{\nerr} \Mcov\ofierr\ofij 
= \sum_{\ierr=1}^{\nerr} \left\{\begin{array}{ll} 
\Mcov\ofierr\ofii
& \hspace{2mm} \mathrm{if} \, \, i=j \myh, \\
\rho\ofij\ofierr\sqrt{\Mcov\ofierr\ofii}\sqrt{\Mcov\ofierr\ofjj}
& \hspace{2mm} \mathrm{if} \, \, i\neq{j} \myh. \\
\end{array}\right.
\label{eq:errsources}
\end{equation}
If the ``nominal'' values 
of the correlations $\rho\ofij\ofierr$ are not assessed rigorously
and it is still conceivable to modify them
to make the combination more ``conservative'',
it is certainly useful to know 
which of these correlations correspond to a high correlation regime
and which of them contribute more significantly to a change in the
combined error. To answer the first question,
one should look at the correlations of those
measurements whose BLUE coefficients are negative
in the ``nominal'' combination.
In particular, one should concentrate on
the correlations $\rho\ofij$ between two measurements $\yexp\ofi$ and $\yexp\ofj$
such that the derivatives $\partial\Info/\partial\rho\ofij$
in Eq.~\ref{eq:d1Iter} are positive.
To answer the second question, we propose to use the 
information derivatives we derived in Sec.~\ref{sec:derivatives}
and look at which correlation would yield 
the largest relative decrease of information $\delta\Info/\Info$
for the same relative rescaling downwards 
$\delta\rho\ofij\ofierr/\rho\ofij\ofierr$ of that correlation.
In other words, we suggest to rank correlations 
by the normalised information derivative
\begin{equation}
\frac{\rho\ofij\ofierr}{\Info}
\left(\frac{\partial\Info}{\partial\rho\ofij\ofierr}\right) 
= -2\Info\lambda\ofi\lambda\ofj\rho\ofij\ofierr\sqrt{\Mcov\ofii\ofierr}\sqrt{\Mcov\ofjj\ofierr} 
= -2\,\frac{\lambda\ofi\lambda\ofj\Mcov\ofij\ofierr}{\syhat^2} \myh,
\label{eq:d1IdSfsByOffDiagByErrSrc}
\end{equation}
where all quantities in the formula
(which is easily derived by extending Eq.~\ref{eq:d1Ibis}) 
are computed at the ``nominal'' values of the correlations.
The correlations 
between measurements $i$ and $j$
and for error source $\ierr$
with the highest positive values 
of the normalized derivative in Eq.~\ref{eq:d1IdSfsByOffDiagByErrSrc} are
those that should be most urgently reassessed.
The quantity in Eq.~\ref{eq:d1IdSfsByOffDiagByErrSrc}
is a dimensionless number: in the simple example presented 
in Sec.~\ref{sec:ri}, for instance,
the value of this normalised derivative
for the correlation between B1 and B2 
(for the single source of uncertainty considered in that example)
is as high as +2.52,
indicating that the combined error would increase by 2.52\% 
for a relative reduction of the correlation
by 1\% of its ``nominal'' value (from 0.87500 to 0.86625).
We will illustrate this in more detail
at the end of this section using a numerical example.

Note that the sums 
of all $(\rho\ofij\ofierr/\Info)(\partial\Info/\partial\rho\ofij\ofierr)$
over all error sources $\ierr$
effectively represent the effect on information
of rescaling the correlations between measurements $i$ and $j$
by the same factor for all error sources,
while their sums over measurements $i$ and $j$
represent the effect 
of rescaling the correlations between all measurements
by the same factor in a given error source $\ierr$.
Likewise, their global sum
over measurements $i$ and $j$ and error sources $\ierr$
represents the effect on information 
of rescaling all correlations by a global factor.
While they lack the granularity to give more useful insight 
about which correlations are most relevant
when trying to make the combination more conservative,
these sums also represent interesting quantities 
to analyse in some situations.
In particular, we will point out in Sec.~\ref{sec:minimization}
that each of these different sums of derivatives 
becomes zero in one of the information minimization procedures
that will be described in that section.

\newcommand{\bluefin}{{\sc BlueFin}}
\newcommand{\vsfs}{\myvec{\sfs}}
\newcommand{\McovNew}{\Mcov'}
\newcommand{\InfoNew}{\Info'}
\newcommand{\lamNew}{\lambda'}

\subsection{Reducing correlations to make them ``more conservative''}

Having proposed a way to identify which ``nominal'' correlations
have not been estimated ``conservatively''
and may need to be reassessed,
we now propose some practical procedures to reduce their values
and try to make the combination more conservative,
when a full and precise reevaluation of these correlations is impossible.
What follows must be understood as simple guidelines to drive 
the work of experimental physicists when 
combining measurements: we propose different methods, but
the applicability of one rather than the other, which  
also implies some level of arbitrariness, would have
to be judged on a case-by-case basis. 

We propose three main solutions to the problem 
of reducing the (large and positive) ``nominal'' values of correlations
to make the combination more conservative: the first is a numerical 
minimization of information with respect to these correlations, 
the second consists in ignoring some of the input measurements, 
and the third one is a prescription that we indicate with
the name of ``onionization'' and
that consists in decreasing the off-diagonal elements in the covariance matrices
so that they are below a specific threshold.
At the end of the section we will present a practical example 
that illustrates the different features of these methods.

\subsubsection{Minimizing information by numerical methods}
\label{sec:minimization}

This approach is based on a multi-dimensional 
minimization of information as a function of rescaling factors 
applied to the ``nominal'' values of correlations.
In the most general case, one would independently rescale
by a scale factor $\sfs\ofierr\ofij$
between 0 and 1 each correlation $\rho\ofierr\ofij$
between the errors on the measurements $\yexp\ofi$ and $\yexp\ofj$
due to the $\ierr^{\mathrm{th}}$ source of uncertainty.
This corresponds to studying the effect on information
of replacing the ``nominal'' covariance matrix $\Mcov\ofierr\ofij$
for each error source
by a modified covariance matrix $(\McovNew)\ofierr\ofij$ given by
\begin{equation}
\Mcov\ofierr\ofij \rightarrow
(\McovNew)\ofierr\ofij =
\left\{
\begin{array}{ll}
\sfs\ofierr\ofij \Mcov\ofierr\ofij
& \mathrm{if} \, i\neq{j} \myh, \\
\Mcov\ofierr\ofii
& \mathrm{if} \, i=j \myh, \\
\end{array}
\right.
\label{eq:sfsgen}
\end{equation}
where
\begin{equation}
0 \leq \sfs\ofierr\ofij  = \sfs\ofierr\ofji \leq 1 \myh.
\label{eq:sfsgencon}
\end{equation}

\begin{sloppypar}
Minimizing information by varying all 
of those $\nerr\!\times\!\nexp(\nexp\!-\!1)/2$ 
scale factors $\sfs\ofierr\ofij$, however, is not an option
because information (i.e. the inverse of the total combined variance)
ultimately depends only on the off-diagonal elements
in the full covariance matrix 
and the minimization would be under-constrained.
We therefore considered three more specific minimization scenarios.
The first scenario, 
which we indicate as ``ByErrSrc'',
consists in rescaling all correlations within each error source 
by the same factor $\sfs\ofierr$,
thus involving $\nerr$ independent scale factors
($\sfs\ofierr\ofij\!=\!\sfs\ofierr$ for every $i$~and $j$).
The second case,
which we indicate as ``ByOffDiagElem'',
consists in rescaling in all error sources
the correlation between $\yexp\ofi$ and $\yexp\ofj$
by the same factor $\sfs\ofij$, 
thus involving $\nexp(\nexp\!-\!1)/2$ independent scale factors
($\sfs\ofierr\ofij\!=\!\sfs\ofij$ for every $\ierr$).
Finally, the simplest case,
which we indicate as ``ByGlobFac'',
consists in rescaling all correlations %
by the same global rescaling factor $\sfs$
(i.e. $\sfs\ofierr\ofij\!=\!\sfs$
for every $i$, $j$ and $\ierr$).
\end{sloppypar}

A software package, called \bluefin\footnote{Best Linear Unbiased Estimate Fisher Information aNalysis
\\
-- \url{https://svnweb.cern.ch/trac/bluefin}},
was specifically prepared to study all these issues.
Within this package, numerical minimizations are performed using 
the {\sc Minuit}~\cite{bib:minuit} libraries through their 
integration in {\sc Root}~\cite{bib:moneta},
imposing the constraints that scale factors remain between 0 and 1.
All scale factors are varied in the minimization, 
except those which are known to have no effect on information
because the information derivatives with respect to them
(which are essentially those presented in Sec.~\ref{sec:derivatives})
are zero both at ``nominal'' and at zero correlations
(i.e. when all scale factors are 1 and 0, respectively).

The ``ByOffDiagElem'' minimization is the most tricky,
as it may trespass into regions 
where the total covariance matrix is not positive definite,
sometimes in an unrecoverable way,
in which case we declare the minimization to have failed.
Even when this minimization does converge to a minimum,
one should also keep in mind that at this point
the partial covariance matrices for the different error sources
may be non positive-definite with negative eigenvalues:
this is clearly a non-physical situation,
which should be used for illustrative purposes only 
and is clearly not suitable for a physics publication.

Not surprisingly,
in the very simple example presented in Sec.~\ref{sec:ri},
where only one off-diagonal correlation is non-zero
and errors are assumed to come from a single source of uncertainty,
these three minimizations all converge to the same result,
where the off-diagonal covariance is reduced to
$\rBoBt\sBo\sBt\!=\!\sBo^2\!=\!16.00$,
which leads to a combination where $\lBt\!=\!0$
and again the less precise measurement B2 is essentially excluded.
In a more general case with several non-zero correlations
and many different sources of uncertainty,
the three minimizations may instead converge to rather different outcomes.
The \bluefin\ software will also be used for the numerical examples shown
at the end of the section.

\subsubsection{Iteratively removing measurements with negative BLUE coefficients}

Having observed many times that choosing ``more conservative'' correlations
may ultimately lead to combinations 
where the BLUE coefficients of one or more measurements
are increased from a negative value to zero,
it is perfectly legitimate to think
of excluding these measurements
from the combination from the very beginning.
If one should choose to adopt this approach,
we suggest to do this iteratively,
by removing first the measurement with the most negative BLUE coefficient,
then performing a new combination and finally iterating 
until only positive BLUE coefficients remain.
This procedure is guaranteed to converge as
the combination of a single measurement
has a single BLUE coefficient equal to 1.
We will present an example later on.

Excluding measurements from a combination may be a very controversial
decision to take. At the same time, if there are negative BLUE 
coefficients and it is impossible to determine precisely the
correlations, this may be the truly conservative and soundest scientific choice,
to avoid the risk of claiming combined results
more accurate than they really are.
Note that excluding a measurement
differs from including it with a rescaled correlation
which gives it a zero BLUE coefficient,
as in the latter case the measurement 
does contribute to the $\chi^2$ for the fit
while in the former case it does not.
If correlations for that measurement 
cannot be precisely assessed, in any case,
even the accuracy of its contribution to the $\chi^2$
with an ad-hoc rescaled correlation is somewhat questionable
and it may be better to simply exclude %
the measurement from the combination altogether.

Note finally that high correlations 
between different measurements in a combination
are not only caused by correlated systematic uncertainties
in the analyses of independent data samples,
but are also expected for statistical uncertainties
when performing different analyses of the same data samples.
In these cases,
where negative BLUE coefficients would be likely 
if two such measurements were combined,
it is already common practice
to only publish the more precise analysis 
and simply use the less precise one as a cross-check.

\newcommand{\aexp}{A}
\newcommand{\bexp}{B}
\newcommand{\cexp}{C}
\newcommand{\dexp}{D}
\newcommand{\xstru}{\ytru}
\newcommand{\xshat}{\yhat}
\newcommand{\xsexp}{\yexp}
\newcommand{\xsa}{\xsexp_\mathrm{A}}
\newcommand{\xsb}{\xsexp_\mathrm{B}}
\newcommand{\xsc}{\xsexp_\mathrm{C}}
\newcommand{\xsd}{\xsexp_\mathrm{D}}
\newcommand{\stat}{UNC}
\newcommand{\bkgd}{BKGD}
\newcommand{\lumi}{LUMI}
\newcommand{\pstat}{\,\mathrm{\scriptstyle(UNC)}}
\newcommand{\pbkgd}{\,\mathrm{\scriptstyle(BKGD)}}
\newcommand{\plumi}{\,\mathrm{\scriptstyle(LUMI)}}

\newcommand{\tableTwo}{
\begin{table*}[t]
\vspace*{1mm}
\begin{center}
\renewcommand{\arraystretch}{1.1}
\begin{tabular}{|lc|ccc|c|c|c|c|}
\hline
\multicolumn{2}{|c|}{Measurements}
& $\sigma_{Unc}$ & $\sigma_{Bkgd}$ & $\sigma_{Lumi}$
& CVW/\%  & IIW/\%  & MIW/\%  & RI/\%  \\
\hline
$\xsa$ &      95.00 $\pm$      17.92 &      10.00 &      10.00 &      11.00 &      60.39 &      50.91 &      34.69 &      48.78 \\
$\xsb$ &     144.00 $\pm$      44.63 &      14.00 &      40.00 &      14.00 &     -11.90 &       8.20 &       8.97 &       9.61 \\
$\xsc$ &     115.00 $\pm$      20.81 &      18.00 &       3.00 &      10.00 &      25.36 &      37.74 &      14.63 &      20.49 \\
$\xsd$ &     122.00 $\pm$      25.00 &      25.00 &          0 &          0 &      26.15 &      26.15 &      26.15 &      21.12 \\
Correlations & --- & --- & --- & --- & --- &     -23.01 & --- & --- \\
\hline
BLUE / Total &     101.30 $\pm$      12.78 &      10.14 &       2.04 &       7.51 &     100.00 &     100.00 &      84.44 &     100.00 \\
\hline
\end{tabular}
\vspace*{2mm}
\caption{Results of the combination 
of $\xsa$, $\xsb$, $\xsc$ and $\xsd$ ($\chi^2$/ndof=4.23/3).
The central value, total error and individual error components
for each input measurement $i$ are listed, followed by
the central value weight CVW$_i$ or $\lambda_i$, 
the intrinsic information weight $\iiw_i$
(also shown for the correlations),
the marginal information weight $\miw_i$, 
the relative importance $\ri_i$.
In the last row,
the BLUE central value and errors
and the sum of all weights in each column are displayed.}
\label{tab:xse}
\vspace*{-6mm}
\end{center}
\end{table*}
}

\subsubsection{The ``onionization'' prescription}

Guided by the remarks we have made so far in this paper,
but without any formal demonstration,
we finally propose a simple rule of thumb 
for trying to modify ``nominal'' correlations 
to make them more ``conservative''.
In the following we will again indicate 
by $\Mcov$ the ``nominal'' covariance
and by $\McovNew$ the modified covariance matrices
(which will only differ in their off-diagonal terms
as we will only modify correlations, keeping variances unchanged).
Our proposal essentially consists 
in generalising Eq.~\ref{eq:rhocons2} to $\nexp$ measurements,
by defining correlations
so that the total modified covariance $(\McovNew)\ofij$ 
between any two measurements 
remains smaller than both individual total variances,
\begin{equation}
\renewcommand{\arraystretch}{1.2}
\biggl\{\begin{array}{ccc}
(\McovNew)\ofij & \leq & (\McovNew)\ofii = \Mcov\ofii = \sigma\ofi^2 \\
(\McovNew)\ofij & \leq & (\McovNew)\ofjj = \Mcov\ofjj = \sigma\ofj^2 \\
\end{array}\biggr. 
\hspace*{2cm}\forall i,j \myh.
\renewcommand{\arraystretch}{1}
\label{eq:rhoconsN}
\end{equation}
Recalling the interpretation of high-correlation regimes
we gave in Sec.~\ref{sec:simpletwomeas},
the idea is to prevent situations 
where one part of the systematic error is 100\% correlated
between different measurements which have a different sensitivity to it,
as their joint action would effectively constrain that effect 
and lead to a reduction of the combined error.
It should be noted that 
this procedure is similar to the ``minimum overlap'' assumption
that was used to estimate the correlation between systematic errors
for different energies or experiments in several QCD studies at LEP,
including, but not limited to, the analysis presented in Ref.~\cite{bib:l3}.

In more detail,
if the full covariance matrix is built 
as the sum of $\nerr$ error sources
as in Eq.~\ref{eq:errsources},
then $\nerr\!\times\!\nexp(\nexp\!-\!1)/2$ correlations 
$(\McovNew)\ofij\ofierr\!/\!\sqrt{(\McovNew)\ofii\ofierr(\McovNew)\ofjj\ofierr}$
need to be separately estimated 
in the $\nerr$ partial covariances $(\McovNew)\ofierr$.
We considered two possible rules of thumb 
to provide conservative estimates 
of the partial covariances satifying Eq.~\ref{eq:rhoconsN}.

The first one consists in requiring that
\begin{equation}
\renewcommand{\arraystretch}{1.2}
\biggl\{\begin{array}{ccc}
(\McovNew)\ofij\ofierr & \leq & (\McovNew)\ofii = \Mcov\ofii = \sigma\ofi^2 \\
(\McovNew)\ofij\ofierr & \leq & (\McovNew)\ofjj = \Mcov\ofjj = \sigma\ofj^2 \\
\end{array}\biggr. 
\hspace*{1cm}\forall i,j, \, \forall \ierr %
\myh,
\renewcommand{\arraystretch}{1}
\label{eq:rhoconsN1}
\end{equation}
i.e. in keeping the $(\McovNew)\ofij\ofierr$ unchanged 
and equal to their ``nominal'' values $\Mcov\ofij\ofierr$
if these satisfy Eq.~\ref{eq:rhoconsN1},
or in reducing them to the upper bounds above otherwise.
This limits the off-diagonal covariance for each error source 
to the maximum allowed for the sum of all such contributions,
but by doing so it does not strictly ensure 
that their sum does not exceed this limit,
hence the resulting correlations may still be overestimated 
with respect to their ``most conservative'' values.

The second rule of thumb consists in requiring that
\begin{equation}
\renewcommand{\arraystretch}{1.2}
\biggl\{\begin{array}{ccc}
(\McovNew)\ofij\ofierr & \leq & (\McovNew)\ofii\ofierr = \Mcov\ofii\ofierr = (\sigma\ofi\ofierr)^2\\
(\McovNew)\ofij\ofierr & \leq & (\McovNew)\ofjj\ofierr = \Mcov\ofjj\ofierr = (\sigma\ofj\ofierr)^2\\
\end{array}\biggr. 
\hspace*{1cm}\forall i,j, \, \forall \ierr %
\myh,
\renewcommand{\arraystretch}{1}
\label{eq:rhoconsN2}
\end{equation}
i.e. again in keeping the $(\McovNew)\ofij\ofierr$ unchanged 
and equal to their ``nominal'' values $\Mcov\ofij\ofierr$
if these satisfy Eq.~\ref{eq:rhoconsN2},
or in reducing them to the upper bounds above otherwise.
This limits the off-diagonal covariance for each error source 
to the maximum allowed when only that error source is considered,
as if there were no others or they were all negligible with respect to it. 
While this second rule of thumb may result in a full covariance matrix
where the off-diagonal covariances are even lower 
than the ``most conservative'' values in Eq.~\ref{eq:rhoconsN},
we believe that this is a more solid procedure,
because it is applied to each error source independently.
In particular, we think that this may guarantee
more ``conservative'' values of the combined BLUE errors 
in the different error sources, and not only for their total.

\begin{sloppypar}
In the following, we will refer to this rule of thumb 
as the ``onionization'' prescription.
In fact, if we consider a set of measurements \{A,B,C,D\ldots\},
ordered so that $\sAcor\ofierr\leq\sBcor\ofierr\leq\sCcor\ofierr\leq\sDcor\ofierr$
for the $\ierr^\mathrm{th}$ source of uncertainty,
this prescription ensures that
the corresponding partial covariance matrix $(\McovNew)\ofierr$
has no unreasonably large off-diagonal element,
but has instead an upper bound 
with a regular layered pattern similar to that of an onion:
\begin{equation}
\left(
\renewcommand{\arraystretch}{1.4}
\begin{array}{c|c|c|c|c}
\multicolumn{1}{c}{(\sAcor\ofierr)^2}&
\multicolumn{1}{c}{(\sAcor\ofierr)^2}&
\multicolumn{1}{c}{(\sAcor\ofierr)^2}&
\multicolumn{1}{c}{(\sAcor\ofierr)^2}&
\ldots\\
\cline{2-5}
(\sAcor\ofierr)^2&
\multicolumn{1}{c}{(\sBcor\ofierr)^2}&
\multicolumn{1}{c}{(\sBcor\ofierr)^2}&
\multicolumn{1}{c}{(\sBcor\ofierr)^2}&
\ldots\\
\cline{3-5}
(\sAcor\ofierr)^2&(\sBcor\ofierr)^2&
\multicolumn{1}{c}{(\sCcor\ofierr)^2}&
\multicolumn{1}{c}{(\sCcor\ofierr)^2}&
\ldots\\
\cline{4-5}
(\sAcor\ofierr)^2&(\sBcor\ofierr)^2&(\sCcor\ofierr)^2&
\multicolumn{1}{c}{(\sDcor\ofierr)^2}&
\ldots\\
\cline{5-5}
\ldots&\ldots&\ldots&\ldots&\ldots\\
\end{array}
\renewcommand{\arraystretch}{1}
\right)\myh.
\end{equation}
Not surprisingly,
in the simple example presented in Sec.~\ref{sec:ri},
the onionization prescription gives the same result 
as the three minimization procedures described above
(i.e. $\rBoBt\sBo\sBt\!=\!\sBo^2\!=\!16.00$),
leading to a combination where $\lBt\!=\!0$
and again the less precise measurement B2 is essentially excluded.
A more complex example is presented below.
\end{sloppypar}

\tableTwo

Note that, in the procedure described above
as well as in its implementation in the \bluefin\ software
that we used to produce the results presented in the next section,
we systematically apply the ``onionization''
of the partial covariance matrix for each source of uncertainty.
In a real combination, it may be more appropriate
to only apply this procedure
to the partial covariance matrices of those sources of uncertainty
for which at least some of the information derivatives 
in Eq.~\ref{eq:d1IdSfsByOffDiagByErrSrc} are positive.
More generally, we stress again that we only propose
this prescription as a rule of thumb,
but no automatic procedure can replace an estimate of correlations 
based on a detailed understanding of the physics processes
responsible for each source of systematic uncertainty.

\subsection{A more complex example}
\label{sec:finalexample}

As an illustration of the tools we presented in this paper, 
we finally present a slightly more complex example representing
the fictitious combination of 
four different cross-section measurements
\aexp, \bexp, \cexp~and \dexp.
For consistency with the notation used so far and 
to avoid any confusion with the use of the symbol $\sigma$ 
to indicate variances,
we refer to the cross-section observable as $\xsexp$,
to its four measurements as $\xsa$, $\xsb$, $\xsc$, $\xsd$,
and to its BLUE as $\xshat$.
Let us assume in the following that the central values 
and errors on the four measurements are given by
\begin{equation}
\left\{
\begin{array}{lcl}
  \xsa / \pb &=&  95.0 \pm 10.0 \pstat \pm 10.0 \pbkgd \pm 11.0 \plumi \myh, \\
  \xsb / \pb &=& 144.0 \pm 14.0 \pstat \pm 40.0 \pbkgd \pm 14.0 \plumi \myh, \\
  \xsc / \pb &=& 115.0 \pm 18.0 \pstat \pm 3.0 \pbkgd \pm 10.0 \plumi \myh, \\
  \xsd / \pb &=& 122.0 \pm 25.0 \pstat \myh, \\
\end{array}
\right.
\label{eq:xse1}
\end{equation}
where \stat\ indicates the uncorrelated errors 
of statistical or systematic origin,
while \bkgd\ and \lumi\ are systematic errors
assumed to be 100\% correlated between the first three experiments,
due for instance to a common background 
and a common luminosity measurement.

\newcommand{\tableThree}{
\begin{table}[b]
\begin{center}
\vspace*{-1mm}
\renewcommand{\arraystretch}{1.1}
\begin{tabular}{|c|ccc|c|}
\hline
 OffDiag \& ErrSrc & {\tiny UNC} & {\tiny BKGD} & {\tiny LUMI} & OffDiag\\
\hline
$\xsb$ / $\xsa$ &  0 &  {    0.352} &  {    0.135} &  {    0.487} \\
$\xsc$ / $\xsa$ &  0 &     -0.056 &     -0.206 &     -0.262 \\
$\xsc$ / $\xsb$ &  0 &      0.044 &  {    0.052} &  {    0.096} \\
$\xsd$ / $\xsa$ &  0 &  0 &  0 &  0 \\
$\xsd$ / $\xsb$ &  0 &  0 &  0 &  0 \\
$\xsd$ / $\xsc$ &  0 &  0 &  0 &  0 \\
\hline
\multirow{2}{*}{ErrSrc} & \multirow{2}{*}{ 0} & \multirow{2}{*}{ {    0.340}} & \multirow{2}{*}{    -0.019} & GlobFact\\
 & & & &  {    0.321} \\
\hline
\end{tabular}
\renewcommand{\arraystretch}{1}
\vspace*{1mm}
\caption{Normalised information derivatives $\rho$/I*dI/d$\rho$ 
for the combination of of $\xsa$, $\xsb$, $\xsc$ and $\xsd$
in the cross-section example, computed at ``nominal'' correlation values.
The last column and last row list information derivatives 
when the same rescaling factor is used for a given off-diagonal element or error source, 
which are equal to the sums of individual derivatives in each row and column, respectively.}
\label{tab:d1Ixse}
\vspace*{-5mm}
\end{center}
\end{table}
}

\newcommand{\tableFour}{
\begin{table*}[t]
\begin{center}
\vspace*{1mm}
\begin{tabular}{|l|l|ccc|c|cccc|}
\hline
Combination & $(\yhat\pm\syhat)/\pb$ 
& \stat & \bkgd  & \lumi & \tiny$\chi^2$/ndof
& $\lambda_\mathrm{A}$ & $\lambda_\mathrm{B}$ 
& $\lambda_\mathrm{C}$ & $\lambda_\mathrm{D}$ \\ 
\hline
``Nominal'' corr. &
101.3 $\pm$ 12.8 & $\pm$ 10.1 & $\pm$ 2.0 & $\pm$ 7.5 & 4.2/3 &
60.4\% & -11.9\% & 25.4\% & 26.1\% \\
\hline
ByGlobFac &
105.2 $\pm$ 13.0 & $\pm$ 9.9 & $\pm$ 4.1 & $\pm$ 7.3 & 3.1/3 &
50.2\% & -5.7\% & 28.6\% & 26.9\% \\
ByErrSrc &
107.3 $\pm$ 13.2 & $\pm$ 9.8 & $\pm$ 4.7 & $\pm$ 7.6 & 2.6/3 &
45.6\% & -1.8\% & 28.2\% & 28.0\% \\
ByOffDiagElem &
108.2 $\pm$ 13.4 & $\pm$ 9.8 & $\pm$ 5.2 & $\pm$ 7.6 & 2.4/3 &
44.1\% & 0.0\% & 27.2\% & 28.7\% \\
No CVWs $<$ 0 &
108.2 $\pm$ 13.4 & $\pm$ 9.8 & $\pm$ 5.2 & $\pm$ 7.6 & 1.3/2 &
44.1\% & --- & 27.2\% & 28.7\% \\
Onionization &
109.2 $\pm$ 13.1 & $\pm$ 9.5 & $\pm$ 4.9 & $\pm$ 7.6 & 2.2/3 &
42.0\% & 2.4\% & 28.3\% & 27.3\% \\
\hline
No corr. &
110.1 $\pm$ 11.5 & $\pm$ 8.8 & $\pm$ 5.0 & $\pm$ 5.6 & 1.6/3 &
41.4\% & 6.7\% & 30.7\% & 21.2\% \\ %
\hline
\end{tabular}
\vspace*{2mm}
\caption{BLUE central values and variances for the cross section example,
with ``nominal'' correlations,
with correlations reduced using the procedures presented in this paper,
as well as with no correlations.}
\label{tab:blue5xse}
\end{center}
\vspace*{-6mm}
\end{table*}
}

The assumption that the background 
is fully correlated between all experiments
may be the result of a detailed analysis,
or a supposedly ``conservative'' assumption
in the absence of more precise correlation estimates.
It is rather unlikely that a more detailed analysis 
would not be performed in a case like this one
--- in particular, in this type of situation,
with such a large difference in the sizes 
of the fully correlated \bkgd\ errors
in the different measurements,
we would recommend to try to split the \bkgd\ systematics 
into its sub-components in the combination ---
but this is clearly an example for illustrative purposes only.

Under the given assumptions,
the results of the BLUE combination 
are those shown in Table~\ref{tab:xse},
where information weights and relative importance are also listed.
There are several comments that can be made about these numbers.
First, if correlation estimates are actually correct,
then the negative BLUE coefficient for $\xsb$
indicates that we are effectively using this measurement
to constrain the background: note, in particular,
the very small final uncertainty on background after the BLUE combination.
Second, $\xsd$ is for all practical purposes 
a measurement independent from $\xsa$, $\xsb$ and $\xsc$:
this is reflected in the fact that 
its intrinsic and marginal information weights
are both equal to its central value weight.
Third, the relative importance of $\xsd$ is clearly underestimated,
while that of $\xsb$ is clearly overestimated,
as it is larger than both its intrinsic and marginal information weights.

\tableThree 

It is also interesting to look in this example
at the normalised information derivatives
described in Sec.~\ref{sec:derivatives}.
These are shown in Table~\ref{tab:d1Ixse}.
The table tells us that the negative BLUE coefficient for $\xsb$
is primarily due to its correlations to $\xsa$,
mainly that between BKGD errors,
but to a lesser extent also that between LUMI errors:
these two information derivatives, in fact,
are both positive and very large.
The correlations between $\xsb$ and $\xsc$
also go in the direction of increasing information,
while those between $\xsa$ and $\xsc$
are in the opposite regime and decrease information.

\tableFour

We now apply to this example the minimization,
negative BLUE coefficient removal 
and onionization prescriptions described in this section.
The results of the combinations performed
after modifying correlations according to these prescriptions
are listed in Table~\ref{tab:blue5xse},
where they are compared to the combination using ``nominal'' correlations 
and another where all correlations have been set to zero.
This table is very interesting because it shows
a wide range of values not only 
for the BLUE combined total error,
but also, and to an even larger extent, 
for the BLUE combined central values
and for the BLUE combined partial errors 
for each source of uncertainty.

\newcommand{\tableFive}{
\begin{table}[b]
\begin{center}
\vspace*{-2mm}
\begin{tabular}{|lccc|}
\hline &&& \\[-1.5ex]
\bkgd\ \hspace*{-1mm} &
$\left(\begin{array}{rrrr}
100. &  400. &  30. & 0. \\
400. & 1600. & 120. & 0. \\
 30. &  120. &   9. & 0. \\
  0. &    0. &   0. & 0. \\
\end{array}\right)$
\hspace*{-2mm} & \!$\rightarrow$\! & \hspace*{-2mm}
$\left(\begin{array}{rrrr}
100. &  100. &   9. & 0. \\
100. & 1600. &   9. & 0. \\
  9. &    9. &   9. & 0. \\
  0. &    0. &   0. & 0. \\
\end{array}\right)$ 
\\[-1.5ex] 
&&& \\ \hline &&& \\[-1.5ex]
\lumi\ \hspace*{-1mm} &
$\left(\begin{array}{rrrr}
121. &  154. & 110. & 0. \\
154. &  196. & 140. & 0. \\
110. &  140. & 100. & 0. \\
  0. &    0. &   0. & 0. \\
\end{array}\right)$
\hspace*{-2mm} & \!$\rightarrow$\! & \hspace*{-2mm}
$\left(\begin{array}{rrrr}
121. &  121. & 100. & 0. \\
121. &  196. & 100. & 0. \\
100. &  100. & 100. & 0. \\
  0. &    0. &   0. & 0. \\
\end{array}\right)$
\\[-1.5ex]
&&& \\ \hline
\end{tabular}
\vspace*{1mm}
\caption{Onionization of the covariance matrices
for the \bkgd\ and \lumi\ error sources
in the cross-section example.
The values are given in pb$^2$.}
\label{tab:covxseonion}
\vspace*{-6mm}
\end{center}
\end{table}
}

The most striking effect, perhaps,
is the fact that all modifications
of the ``nominal'' correlations 
to make them ``more conservative''
lead to significant central value shifts
(i.e. possibly to biased combined estimates)
and to much larger combined \bkgd\ systematics,
in spite of relatively small increases 
in the total combined errors.
In particular, it is somewhat counter-intuitive
that the combined uncorrelated error decreases
when reducing correlations,
while the combined systematic errors increase:
this is likely to be another feature
of the high-correlation regime
characterizing the ``nominal'' correlations of this example.
We stress that, in real %
situations, it is important to analyse this type of effects,
and not only the effect on the total combined error,
when testing different estimates of correlations.
This is especially important if one keeps in mind
that the results of BLUE combinations 
are generally meant to be further combined with other results
(e.g. the combined top masses from LHC 
and the combined top mass from Tevatron 
will eventually be combined).

It is not too surprising, conversely,
that the effects on combined \bkgd\ systematics
are much larger than those on the combined \lumi\ systematics.
This could be guessed by remembering
that normalised information derivatives
are much larger for the former than for the latter.

It is also not surprising
that the ``ByOffDiagElem'' minimization
gives essentially the same results 
(except for the $\chi^2$ value)
that are found when excluding the measurements 
with negative BLUE coefficients.
By construction, in fact, this is the only one 
of the three minimizations which almost always guarantees
that BLUE coefficients which were initially negative
end up equal to zero after the minimization:
if the minimum is a local minimum,
some of the derivatives in Eq.~\ref{eq:d1Iter},
which are directly proportional to the BLUE coefficients,
must eventually be zero.

\tableFive

\begin{table}[t]
\vspace*{1mm}
\begin{center}
\begin{tabular}{|lc|}
\hline & \\[-1.5ex]
\hspace*{1mm} ``Nominal'' corr. &
$\left(\begin{array}{rrrr}
321. &  554. & 140. &   0. \\
554. & 1992. & 260. &   0. \\
140. &  260. & 433. &   0. \\
  0. &    0. &   0. & 625. \\
\end{array}\right)$ \hspace*{1mm} \\[-1.5ex]
& \\ \hline & \\[-1.5ex]
\hspace*{1mm} ByGlobFac &
$\left(\begin{array}{rrrr}
321. &  442. & 112. &   0. \\
442. & 1992. & 208. &   0. \\
112. &  208. & 433. &   0. \\
  0. &    0. &   0. & 625. \\
\end{array}\right)$ \hspace*{1mm} \\[-1.5ex]
& \\ \hline & \\[-1.5ex]
\hspace*{1mm} ByErrSrc &
$\left(\begin{array}{rrrr}
321. &  341. & 124. &   0. \\
341. & 1992. & 196. &   0. \\
124. &  196. & 433. &   0. \\
  0. &    0. &   0. & 625. \\
\end{array}\right)$ \hspace*{1mm} \\[-1.5ex]
& \\ \hline & \\[-1.5ex]
\hspace*{1mm} ByOffDiagElem &
$\left(\begin{array}{rrrr}
321. &  272. & 140. &   0. \\
272. & 1992. & 219. &   0. \\
140. &  219. & 433. &   0. \\
  0. &    0. &   0. & 625. \\
\end{array}\right)$ \hspace*{1mm} \\[-1.5ex]
& \\ \hline & \\[-1.5ex]
\hspace*{1mm} Onionization &
$\left(\begin{array}{rrrr}
321. &  221. & 109. &   0. \\
221. & 1992. & 109. &   0. \\
109. &  109. & 433. &   0. \\
  0. &    0. &   0. & 625. \\
\end{array}\right)$ \hspace*{1mm} \\[-1.5ex]
& \\ \hline
\end{tabular}
\vspace*{1mm}
\caption{Modified input covariances for the four measurements 
in the cross-section example, when reducing correlations
according to the procedures described in this paper.
The values are given in pb$^2$.}
\label{tab:covxse}
\vspace*{-6mm}
\end{center}
\end{table}

\begin{sloppypar}
Note also that the onionization prescription
leads to the only combination
where the BLUE coefficient for measurement $\xsb$ becomes strictly positive.
As mentioned earlier, this may be a consequence 
of the fact that this prescription may reduce correlations 
even more than their ``most conservative'' values,
trespassing well into the low-correlation regime.
In this respect, it is interesting
to have a look at the effect of onionization
on the partial covariance matrices,
and more generally at the effect
on the total covariance matrices
of all procedures presented in this section:
these are shown in Tables~\ref{tab:covxseonion}
and~\ref{tab:covxse}, respectively.
\end{sloppypar}

\newcommand{\postponeOne}{
It should finally be added that the total covariance matrix
derived from the onionization prescription
is used as the starting point of the ``ByOffDiagElem'' minimization 
in the \bluefin\ software, as we have found this 
to improve the efficiency of the minimization procedure.
As an additional cross-check of the onionization prescription,
we also tested a fourth type of minimization,
where information is independently minimized 
for each source of uncertainty as if this was the only one,
varying each time only the correlations in that error source
(after removing those measurements not affected by it
and slightly reducing the allowed correlation ranges 
to keep the partial covariance positive definite).
The preliminary results of this test
(not included in Table~\ref{tab:blue5xse})
indicate that these minimizations do not seem 
to significantly move partial covariances or the final result away
from those obtained through the onionization prescription,
which are used as a starting point also in this case.
}

In particular, note in Table~\ref{tab:covxseonion}
that the onionization procedure
(but the same is true for minimizations)
affects correlations for the \bkgd\ and \lumi\ error sources
in exactly the same way without distinctions.
If this was a real combination, instead,
one would most likely keep the \lumi\ correlation unchanged
(because a common luminosity measurement would indeed
result in a 100\% correlation between $\xsa$, $\xsb$ and $\xsc$,
and these three measurements together could even help to constrain the error on it),
concentrating instead on the re-assessment 
of the \bkgd\ correlation alone
(because the initial ``nominal'' estimate of 100\% correlation
is neither conservative nor realistic
in the presence of different sensitivities to differential distributions).

\postponeOne

We conclude this section by reminding
that the prescriptions presented here
are only empirical recipes 
that assume no prior knowledge of the physics involved
and, for this reason, can never represent valid substitutes
for a careful quantitative analysis of correlations
using real or simulated data.
A precise estimate of correlations is important
in general, but absolutely necessary in high correlation regimes,
where it may be as important as a precise assessment of measurement errors themselves.

\section{Conclusions}
\label{sec:conclusions}

Combining many correlated measurements 
is a fundamental and unavoidable step 
in the scientific process 
to improve our knowledge about a physical quantity.
In this paper, we recalled the relevance 
of the concept of Fisher information
to quantify and better understand this knowledge.
We stressed that it is extremely important 
to understand how the information available from several measurements 
is effectively used in their combination,
not only because this allows a fairer recognition of their relative merit
in their contribution to the knowledge about the unknown parameter,
but especially because this makes it possible
to produce a more robust scientific result
by critically reassessing the assumptions made in the combination.

In this context, we described how
the correlations between the different measurements 
play a critical role in their combination.
We demonstrated, in particular,
that the presence of negative coefficients
in the BLUE weighted average of any number of measurements is a sign %
of a ``high-correlation regime'',
where the effect of increasing correlations
is that of reducing the error on the combined result.
We showed that, in this regime,
a large contribution to the combined knowledge about the parameter
comes from the joint impact of several measurements through their correlation
and we argued, as a consequence, 
that the merit for this particular contribution to information 
cannot be claimed by any single measurement individually.
In particular, we presented our objections
to the standard practice of presenting 
the ``relative importances'' of different measurements
based on the absolute values of their BLUE coefficients,
and we proposed the use of  
(``intrinsic'' and ``marginal'') ``information weights'' instead.

In the second part of the paper,
we questioned under which circumstances
assuming systematic errors as fully correlated 
can be considered a ``conservative'' procedure.
We proposed the use of information derivatives
with respect to inter-measurement correlations
as a tool to identify those ``nominal'' correlations 
for which this assumption is wrong
and a more careful evaluation is necessary.
We also suggested a few procedures
for trying to make a combination more ``conservative''
when a precise estimate of correlations is simply impossible.

We should finally note that BLUE combinations
are not the only way to combine different measurements,
but they are actually the simplest to understand
when combinations are performed under the most favorable assumptions
that measurements are multivariate Gaussian distributed
with covariances known a priori,
as in this case all relevant quantities 
become easily calculable by matrix algebra.
We therefore stress that,
while the results in this paper were obtained 
under these assumptions and using the BLUE technique,
large positive correlations 
are guaranteed to have a big impact,
and should be watched out for,
also in combinations performed with other methods
or under other assumptions.

\section*{Acknowledgements}
This work has been inspired by many discussions,
during private and public meetings, 
on the need for critically reviewing 
the assumptions about correlations and the meaning of ``weights'',
when combining several measurements
in the presence of high correlations between them.
It would be difficult to mention and acknowledge 
all those colleagues who have hinted us towards the right direction 
and with whom we have had very fruitful discussions. 
We are particularly grateful to the members of the TopLHCWG 
and to the ATLAS and CMS members who have helped in the reviews 
of the recent top mass combinations at the LHC.

We would also like to thank our colleagues who have sent us 
comments about the first two public versions of this paper.
In particular, it is a pleasure to thank Louis Lyons
for his extensive feedback and his very useful suggestions.
We are also grateful to the EPJC referees for their detailed
and insightful comments, as well as for making us aware 
of the research presented in Ref.~\cite{bib:cox}.

Finally, A.V. would like to thank the management
of the CERN IT-ES and IT-SDC groups for allowing him 
the flexibility to work on this research
alongside his other committments 
in computing support for the LHC experiments.

\newcommand{\NIMA}[3] {Nucl.\ Instr.\ Meth.\ {A#1} (#2) #3}

\end{document}